\documentclass{article}

\usepackage{arxiv}

\usepackage[utf8]{inputenc} % allow utf-8 input
\usepackage[T1]{fontenc}    % use 8-bit T1 fonts
\usepackage{hyperref}       % hyperlinks
\usepackage{url}            % simple URL typesetting
\usepackage{booktabs}       % professional-quality tables
\usepackage{amsfonts}       % blackboard math symbols
\usepackage{nicefrac}       % compact symbols for 1/2, etc.
\usepackage{microtype}      % microtypography
\usepackage{lipsum}
\usepackage{graphicx}
\graphicspath{ {./images/} }

\usepackage{booktabs} 
\usepackage{subcaption}
\usepackage{mathtools}
\usepackage{multicol}
\usepackage{color}
\usepackage{colortbl}
\usepackage{tabularx,colortbl}
\usepackage{caption}
\usepackage{multirow}
\usepackage{multicol}
\usepackage{enumitem}
\usepackage{ltablex}
\usepackage{makecell}
\usepackage{lscape}
\usepackage{xcolor}
\usepackage{varwidth}
\usepackage{caption}
\usepackage[font=small,skip=5pt,belowskip=-1pt]{caption}
\usepackage{url}
\usepackage{graphics}
\usepackage{balance}
\usepackage{array}
\usepackage{appendix}
\usepackage{array}

\usepackage{geometry}
\usepackage{amsmath}
\usepackage{ragged2e}

\makeatletter
\newcommand*{\compress}{\@minipagetrue}
\makeatother
\newcolumntype{Q}{ >{\compress\itemize}X<{\enditemize}}

\newcolumntype{R}{>{\raggedright\arraybackslash}X}

\newcolumntype{P}[1]{>{\centering\arraybackslash}p{#1}}
\newcolumntype{Y}{>{\centering\arraybackslash}X}% for centering text horizontally
\definecolor{LightGrey}{rgb}{0.93,0.93,0.93}

\newcommand\boldGrey[1]{\cellcolor{LightGrey}{\textbf{#1}}}
\definecolor{mygrey}{gray}{0.8}

\newcolumntype{s}{>{\hsize=.65\hsize}X}

\title{Understanding the ``Pathway'' Towards a Searcher's Learning Objective}

\author{
 Kelsey Urgo \\
  School of Information and Library Science\\University of North Carolina at Chapel Hill\\
  Chapel Hill, NC \\
  \texttt{kurgo@unc.edu} \\
   \And
 Jaime Arguello \\
  School of Information and Library Science\\University of North Carolina at Chapel Hill\\
  Chapel Hill, NC \\
  \texttt{jarguello@unc.edu} \\
}

\begin{document}
\maketitle
\begin{abstract}
Search systems are often used to support learning-oriented goals. This trend has given rise to the ``search-as-learning'' movement, which proposes that search systems should be designed to support learning. To this end, an important research question is: How does a searcher's \emph{type} of learning objective influence their trajectory (or \emph{pathway}) towards that objective? We report on a lab study ($N=36$) in which participants gathered information to meet a specific type of learning objective. To characterize learning objectives \emph{and pathways}, we leveraged Anderson and Krathwohl's (A\&K's) taxonomy~\cite{anderson2001taxonomy}. A\&K's taxonomy situates learning objectives at the intersection of two orthogonal dimensions: (1) \emph{cognitive process} (remember, understand, apply, analyze, evaluate, and create) and (2) \emph{knowledge type} (factual, conceptual, procedural, and metacognitive knowledge). Participants completed learning-oriented search tasks that varied along three cognitive processes (apply, evaluate, create) and three knowledge types (factual, conceptual, procedural knowledge). A \emph{pathway} is defined as a sequence of \emph{learning instances} (e.g., subgoals) that were also each classified into cells from A\&K's taxonomy. Our study used a think-aloud protocol, and pathways were generated through a qualitative analysis of participants' think-aloud comments and recorded screen activities. We investigate three research questions. First, in RQ1, we study the impact of the learning objective on pathway characteristics (e.g., pathway length). Second, in RQ2, we study the impact of the learning objective on the types of A\&K cells traversed along the pathway. Third, in RQ3, we study common and uncommon \emph{transitions} between A\&K cells along pathways conditioned on the knowledge type of the objective. We discuss implications of our results for designing search systems to support learning.
\end{abstract}

% keywords can be removed

\section{Introduction}

People often search for information in order to learn something new. While current search systems are effective in helping users complete simple look-up tasks (e.g., navigational or fact-finding tasks), they provide less support for users working on complex tasks that involve learning. In recent years, the ``search-as-learning'' research community has argued that search systems should be better designed to support learning. Recent summits have taken place to develop research agendas in the area of search-as-learning~\cite{allan_frontiers_2012,collins-thompson_search_2017}.  Participants at these summits proposed that future research should focus on: (1) understanding the contexts in which people search for information in order to learn, (2) understanding the cognitive biases promoted by existing search systems, (3) understanding search as a learning process, and (4) developing search interfaces and tools that encourage and support learning~\cite{allan_frontiers_2012,collins-thompson_search_2017}. Our research in this paper focuses on understanding search as a learning process.  Additionally, we discuss how our results have implications for designing novel search tools to encourage and support learning.

Prior studies in the area of search-as-learning have investigated a wide range of research questions.  Many studies have investigated how different factors can influence learning during search.  Specifically, studies have investigated characteristics of the individual searcher~\cite{obrien_role_2020,pardi_role_2020,roy_exploring_2020,willoughby_fast_2009}, characteristics of the search task~\cite{ghosh_searching_2018,kalyani_understanding_2019,liu_examining_2013,liu_investigation_2019}, and characteristics of the search system~\cite{freund_effects_2016,camara_searching_2021,chi_exploring_2016,demaree_influence_2020,heilman_language_2006,heilman_personalization_2010,hersh_towards_1995,kammerer_signpost_2009,palani_active_2021,qiu_towards_2020,roy_note_2021,syed_improving_2020,syed_optimizing_2017,weingart_retrieval_2016,wilson_backward_2008,xu_how_2020}.  Additionally, studies have investigated the relation between specific search behaviors and learning outcomes~\cite{abualsaud_learning_2017,bhattacharya_measuring_2019,chi_exploring_2016,collins-thompson_assessing_2016,gadiraju_analyzing_2018,lei_effect_2015,liu_how_2018,lu_personalized_2017,palani_active_2021,xu_how_2020,yu_predicting_2018}. 

When people search to learn, they typically have a specific \emph{learning objective} in mind---``I need to find information that enables me to do <learning objective>.''  Our research in this paper investigates an important question that has not been directly addressed in prior work: 

\vspace{.1in}
\begin{quote}
How does the type of \emph{learning objective} that a searcher is aiming to accomplish influence their trajectory or \emph{pathway} towards that objective?  In other words, how do searchers decompose a specific objective into a sequence of subgoals or learning instances?
\end{quote}
\vspace{.1in}

To address this question, we conducted a lab study in which participants ($N=36$) completed search tasks with different types of learning objectives. To gain insights about participants' pathways  (i.e., sequences of subgoals) towards an objective, the study used a think-aloud protocol.  To manipulate learning objectives \emph{and} to characterize pathways towards an objective, we leveraged Anderson and Krathwohl's taxonomy of learning (referred to as A\&K's taxonomy)~\cite{anderson2001taxonomy}.

\subsection{A\&K's Taxonomy} 

In the field of education, A\&K's taxonomy was developed to help educators more precisely define learning objectives for students~\cite{anderson2001taxonomy}.  Additionally, it was developed to help educators \emph{align} instructional exercises and assessments with the target learning objective. As illustrated in Table~\ref{tab:pathway}, A\&K's taxonomy situates learning objectives at the intersection of two orthogonal dimensions: (1) cognitive process and (2) knowledge type. Anderson and Krathwohl~\cite{anderson2001taxonomy} argued that learning objectives can be viewed as a combination of a ``verb'' and a ``noun'' (e.g., recall factual knowledge). In this respect, the cognitive process defines the ``verb'' and the knowledge type defines the ``noun'' of the learning objective.

The \emph{cognitive process} dimension defines the types of cognitive activities associated with the learning objective. In other words, it defines the types of mental activities learners should be able to perform once the objective is met. Cognitive processes range from simple to complex: remember, understand, apply, analyze, evaluate, and create. If a \emph{remember} objective is met, it means that the learner will be able to recall or regurgitate information verbatim. If an \emph{understand} objective is met, it means that the learner will be able to explain information in their own words or illustrate examples of a construct. If an \emph{apply} objective is met, it means that the learner will be able to execute a process or use the acquired knowledge in a new scenario. If an \emph{analyze} objective is met, it means that the learner will be able to explain relations between elements (e.g., similarities and differences). If an \emph{evaluate} objective is met, it means that the learner will be able to critique or prioritize elements. Finally, if a \emph{create} objective is met, it means that the learner will be able to generate a new solution to a problem or organize information using a novel representation.

The \emph{knowledge type} dimension defines the type of knowledge associated with the objective. A\&K's taxonomy defines four types of knowledge: factual, conceptual, procedural, and metacognitive knowledge. The first three knowledge types relate to external knowledge about the world: \emph{factual knowledge} relates to self-contained, objective bits of information; \emph{conceptual knowledge} relates to concepts, categories, theories, principles, schemas, and models; and \emph{procedural knowledge} relates to knowledge about how to perform a task. Conversely, \emph{metacognitive knowledge} looks inward, and relates to knowledge about one's own cognition or cognition in general.

\vspace{-.2cm}
\subsection{Characterizing Pathways} 
\vspace{-.1cm}
Our goal is to understand the pathways followed by searchers towards a specific learning objective. A \emph{pathway} is defined as a sequence of \emph{learning instances} towards an objective. A \emph{learning instance} is defined as a point during the search and learning process in which the searcher either: (1) sets forth a new learning-oriented subgoal or (2) serendipitously learns something new and useful towards achieving the objective. The A\&K taxonomy can be leveraged to classify learning objectives and learning instances along the pathway to the objective.

To better understand the concept of a pathway, consider the example in Table~\ref{tab:pathway}. Imagine a searcher who wants to determine which sorting algorithm is more efficient: quicksort or mergesort. This learning objective involves making a judgement (i.e., evaluate) between two algorithms (i.e., procedural knowledge).  Therefore, based on A\&K's taxonomy, the objective can be classified as evaluate/procedural (gray cell in Table \ref{tab:pathway}). Given this objective, a searcher may follow the pathway below, which includes learning instances (LIs) that are either planned or unplanned.

\begin{table}[t]
\caption{A\&K's two-dimensional taxonomy and example pathway towards an evaluate/procedural learning objective (gray cell). LI$_i$ denotes the $i$th learning instance along the pathway.}\label{tab:pathway}
\begin{tabular}{|l|l|l|l|l|l|l|}
\hline
 \multirow{2}{*}{Knowledge Type} & \multicolumn{6}{c|}{Cognitive Process} \\
 \cline{2-7}
 & Remember & Understand & Apply & Analyze & Evaluate & Create \\\hline
Factual & & & & & & \\\hline
Conceptual & & LI$_2$ & & & & \\\hline
Procedural & & LI$_1$, LI$_3$, LI$_4$ & LI$_5$ & LI$_6$, LI$_7$ & \cellcolor[HTML]{D9D9D9} LI$_8$ & \\\hline
Metacognitive & & & & & & \\\hline
\end{tabular}
\vspace{-.4cm}
\end{table}

\begin{itemize}
\item \textbf{LI}$_{1}$ \textbf{(understand/procedural):} First, the searcher may look for and review an example of mergesort to understand the steps.
\item \textbf{LI}$_{2}$ \textbf{(understand/conceptual):} Second, while pursuing LI$_{1}$, the searcher may encounter the concept of ``divide and conquer'' and look for a definition to understand this concept.
\item \textbf{LI}$_{3}$ \textbf{(understand/procedural):} Third, to deepen their understanding of mergesort, the searcher may review its pseudocode (i.e., a different representation of the procedure).
\item \textbf{LI}$_{4}$ \textbf{(understand/procedural):} Fourth, the searcher may look for and review an example of quicksort to understand the steps. 
\item \textbf{LI}$_{5}$ \textbf{(apply/procedural):} Fifth, to deepen their understanding of how the algorithm works, the searcher may decide to sort a list of numbers using quicksort.
\item \textbf{LI}$_{6}$ \textbf{(analyze/procedural):} Sixth, upon realizing that mergesort and quicksort are \emph{both} ``divide and conquer'' algorithms, the searcher may review an article that explains how both algorithms use ``divide and conquer''.
\item \textbf{LI}$_{7}$ \textbf{(analyze/procedural):}  Seventh, the searcher may read an article that explains the similarities and differences between mergesort and quicksort.
\item \textbf{LI}$_{8}$ \textbf{(evaluate/conceptual):} Finally, the searcher may read an article that explains why mergesort is better than quicksort for large arrays, and may use this information as a rationale to judge that mergesort is more efficient in real-world scenarios.
\end{itemize}

As illustrated in the example above, A\&K's taxonomy can be leveraged to categorize learning objectives and learning instances traversed along the pathway towards the objective.  Learning objectives and instances along the pathway can all be assigned to cells in A\&K's taxonomy.

\vspace{-.2cm}
\subsection{Study Overview and Research Questions}
\vspace{-.1cm}
Participants in the study completed three search tasks with learning objectives situated at the intersection of \emph{three} cognitive processes (apply, evaluate, create) and \emph{three} knowledge types (factual, conceptual, procedural). To analyze the pathways taken by participants towards an objective, we performed a qualitative analysis of search sessions based on participants' think-aloud comments and recorded search and note-taking activities. The study investigated three main research questions (RQ1-RQ3), which all centered on the pathways participants followed towards an objective.

A preliminary analysis of pathways found that participants primarily stayed within the \emph{same} knowledge type as the learning objective they were asked to accomplish. As illustrated in the example in Table~\ref{tab:pathway}, pathways towards a procedural objective mostly involved learning instances focused on procedural knowledge. Therefore, our analysis of pathways focused exclusively on the \emph{cognitive processes} traversed along the pathways.

We investigate the following research questions:

\begin{itemize}
\item \textbf{RQ1:}  What is the effect of the learning objective (i.e., cognitive process and knowledge type)  on the characteristics of pathways towards the objective?
\item \textbf{RQ2:} What is the effect of the learning objective (i.e., cognitive process and knowledge type)  on the cognitive processes traversed along pathways towards the objective?
\item \textbf{RQ3:} What are common and uncommon \emph{transitions} between cognitive processes traversed along pathways towards the objective?
\end{itemize}

In RQ1, we investigate the effects of the learning objective on two pathway characteristics: (1) pathway length (i.e., number of learning instances) and (2) pathway diversity (i.e., number of distinct cognitive processes traversed). We investigate these differences from two perspectives.  First, we explore pathway differences by conditioning on the cognitive process of the objective.  For example, are pathways longer or more diverse when the objective is to create versus apply?  Second, we explore pathway differences by conditioning on the knowledge type of the objective. For example, are pathways longer or more diverse during procedural versus factual objectives?

In RQ2, we investigate the effects of the learning objective on the cognitive processes traversed along the pathways. As in RQ1, we investigate these differences from two perspectives.  First, we explore pathway differences by conditioning on the cognitive process of the objective.  For example, do pathways involve more analyze learning instances when the objective is to evaluate versus apply? Second, we explore pathway differences by conditioning on the knowledge type of the objective. For example, do pathways involve more analyze learning instances when the objective involves conceptual versus factual knowledge?

In RQ3, we investigate the types of cognitive process \emph{transitions} along pathways towards an objective.  Again, we investigate this question from two perspectives.  First, we consider common and  uncommon transitions \emph{irrespective of the objective}. For example, regardless of the objective, are searchers more likely to transition from simple to complex processes  (e.g., understand to analyze) or more likely to transition from complex to simple processes (e.g., analyze to understand)?

Second, we consider common and uncommon transitions by \emph{conditioning on the objective}. As described below (Section~\ref{subsec:ictir2020}), in a previous paper~\cite{UrgoICTIR2020}, we reported on results derived from the same lab study. Specifically, we reported on the effects of the learning objective (i.e., cognitive process and knowledge type) on participants' pre- and post-task perceptions and search behaviors. Our results found that the knowledge type of the objective had a \emph{much stronger} effect than the cognitive process of the objective. Based on these results, we decided to focus our RQ3 analysis by conditioning only on the knowledge type of the objective.  For example, are searchers more likely to transition from simple to complex cognitive processes during objectives involving procedural versus conceptual knowledge?

\vspace{-.2cm}
\subsection{Extension of Prior Work}\label{subsec:ictir2020}
\vspace{-.1cm}
Our research in this paper is an extension of our own prior work. In Urgo et al.~\cite{UrgoICTIR2020}, we reported on results from the same study.  Specifically, we reported on the effects of the learning objective (i.e., cognitive process and knowledge type) on: (1) pre-task perceptions, (2) post-task perceptions, and (3) search behaviors.  Interestingly, the objective's cognitive process (apply vs.~evaluate vs. create) had \emph{no significant effects}.  Conversely, the objective's knowledge type (factual vs.~conceptual vs.~procedural) had several significant effects.  First, factual objectives were perceived to require \emph{less} cognitive activity along processes more complex than `remembering'.  Second, conceptual objectives were perceived to require more `understanding' and `analyzing'.  Finally, procedural objectives were perceived to require more `applying', `evaluating' and `creating'. In this paper, we extend this prior work by analyzing the pathways taken by participants towards a specific objective.  In Section~\ref{sec:discussion}, we describe how our results relate with those reported in Urgo et al.~\cite{UrgoICTIR2020}.

\section{Related Work}\label{sec:related_work}
\vspace{-.1cm}
Our research builds on two areas of prior work: (1) search-as-learning and (2) understanding how task characteristics can influence search behaviors and outcomes.

\vspace{-.2cm}
\subsection{Search-as-Learning}
\vspace{-.1cm}

Studies in the area of search-as-learning have investigated a wide range of research questions. Some studies have investigated factors that influence learning during search. Specifically, studies have focused on how learning is impacted by characteristics of the searcher, search task, and search system. Additionally, studies have investigated how learning outcomes are related to specific search behaviors. In the following sections, we review key insights gained from these prior studies in search-as-learning. We focus primarily on key takeaways with respect to learning outcomes.\footnote{In our review, we focus primarily on key takeaways with respect to learning outcomes.  As a side note, prior studies have used a \emph{wide} variety of methods to measure learning. Some studies have measured learning by administering pre- and post-tests with predefined correct answers, including: (1) true-or-false~\cite{freund_effects_2016,gadiraju_analyzing_2018,qiu_towards_2020,kalyani_understanding_2019,yu_predicting_2018,nelson_little_2009}, (2) multiple-choice~\cite{freund_effects_2016,kalyani_understanding_2019,heilman_language_2006,heilman_personalization_2010,syed_optimizing_2017,weingart_retrieval_2016,davies_self-regulated_2013} and (3) short-answer tests~\cite{hersh_towards_1995,collins-thompson_assessing_2016,roy_exploring_2020,abualsaud_learning_2017,davies_self-regulated_2013,camara_searching_2021,roy_note_2021}. Other studies have asked participants to complete more \emph{open-ended} exercises. Specifically, studies have measured learning by asking participants to: (1) list relevant key phrases and facts~\cite{bhattacharya_measuring_2019, kammerer_signpost_2009}; (2) create visual representations of a domain~\cite{liu_investigation_2019}; (3) enumerate arguments for and against a specific proposition~\cite{demaree_influence_2020}; and (4) summarize their knowledge of a topic~\cite{collins-thompson_assessing_2016,abualsaud_learning_2017,kalyani_understanding_2019,lei_effect_2015,obrien_role_2020,palani_active_2021,pardi_role_2020,salmeron_using_2020,davies_self-regulated_2013,willoughby_fast_2009,liu_how_2018}. To assess learning from open-ended responses, studies have adopted grading strategies that involve: (1) counting relevant concepts or facts~\cite{bhattacharya_measuring_2019,willoughby_fast_2009,collins-thompson_assessing_2016,abualsaud_learning_2017,palani_active_2021,kammerer_signpost_2009}; (2) counting relevant pro/con arguments~\cite{demaree_influence_2020}; and (3) counting statements that show evidence of generalization or critical thinking~\cite{collins-thompson_assessing_2016,abualsaud_learning_2017,obrien_role_2020,palani_active_2021,liu_how_2018,salmeron_using_2020}. Finally, studies have also considered self-reported \emph{perceptions of learning}~\cite{collins-thompson_assessing_2016,ghosh_searching_2018,liu_investigation_2019,kammerer_signpost_2009,heilman_personalization_2010,freund_effects_2016} and \emph{behavioral measures} that are \emph{assumed} to provide evidence of learning~\cite{chi_exploring_2016}.}

\vspace{-.2cm}
\subsubsection{\textbf{The Effects of User Characteristics on Learning}}

Several studies have investigated the effects of domain knowledge on learning during search~\cite{obrien_role_2020,willoughby_fast_2009,roy_exploring_2020}. O'Brien et al.~\cite{obrien_role_2020} measured learning by asking participants to produce knowledge summaries before and after completing three search tasks on the same general topic. Compared to domain experts, novices had slightly greater improvements in their summary scores. One explanation is that novices uncovered more \emph{new} information while searching. Willoughby et al.~\cite{willoughby_fast_2009} asked participants to produce knowledge summaries on domains where they had high and low prior knowledge. Additionally, one group of participants was instructed to search for 30 minutes before producing their summaries and another group produced summaries without searching. Participants in the \emph{search condition} produced summaries with more accurate facts.  Interestingly, however, this effect was only found when participants had high prior knowledge. The authors hypothesized that participants with higher prior knowledge were able to search more effectively. Roy et al.~\cite{roy_exploring_2020} investigated the role of domain knowledge on learning \emph{during} the search session. To this end, participants completed quick vocabulary learning assessments at regular intervals during the session. Prior knowledge influenced \emph{when} participants had the greatest knowledge gains---towards the start of the session for participants with low prior knowledge and towards the end of the session for participants with high prior knowledge.

To summarize, results suggest a complex relationship between domain knowledge and learning during search. Specifically, the effects of domain knowledge are likely to depend on \emph{other} factors, such as the complexity of the task domain. For example, within simple domains, novices may learn \emph{more} because they simply start with less prior knowledge. Conversely, within complex domains, novices may learn \emph{less} because they lack the prerequisite knowledge to search effectively.

Beyond domain knowledge, prior work has also considered the impact of individual abilities on learning during search. Pardi et al.~\cite{pardi_role_2020} considered the impact of working memory capacity and reading comprehension ability. Learning was measured based on the number of relevant concepts included in knowledge summaries produced by participants before and after searching. Both abilities had a positive effect on learning.

\vspace{-.2cm}
\subsubsection{\textbf{The Effects of Task Characteristics on Learning}}

Several studies have investigated how task characteristics influence learning during search. Similar to our work, studies have leveraged the A\&K taxonomy to study learning-oriented search tasks involving different cognitive processes. Ghosh et al.~\cite{ghosh_searching_2018} had participants complete tasks associated with the cognitive processes of understand, apply, analyze, and evaluate. Participants self-reported significant knowledge gains across all tasks. Additionally, participants were asked to select `action verbs' describing their mental activities during each task. Participants selected different action verbs for each task type---`define' for remember, `demonstrate' for apply, and `relate' for analyze and evaluate tasks. Kalyani and Gadiraju~\cite{kalyani_understanding_2019} had participants complete tasks associated with all six cognitive processes from the A\&K taxonomy. Learning was measured using closed-ended tests for simple tasks and open-ended tests for complex tasks. Participants had lower knowledge gains for complex tasks (i.e., apply < evaluate). Liu et al.~\cite{liu_investigation_2019} had participants complete two tasks of varying cognitive complexity: a \emph{receptive} (i.e., remember or understand) task and a \emph{critical} (i.e., evaluate) task. To measure learning, participants constructed mind maps (i.e., graphical domain representations) before each task, and modified their mind maps throughout the search session. During receptive tasks, participants made structural changes to their mind maps throughout the whole session. Conversely, during critical tasks, participants made more structural changes towards the end of the session.

Beyond task complexity, research has also studied learning during multi-session search. Liu et al.~\cite{liu_examining_2013} had participants complete three subtasks on the same general topic. In the dependent subtask condition, all three subtasks built on each other. Conversely, in the parallel subtask condition, the three subtasks were largely independent (i.e., could be hypothetically done in any order). To measure learning, participants rated their familiarity with the general topic after each subtask. As expected, participants reported greater topic familiarity after each subtask. Interestingly, however, this increase in topic familiarity plateaued faster in the parallel (vs.~dependent) subtask condition, suggesting that searchers benefit from subtasks that build on each other.

\vspace{-.2cm}
\subsubsection{\textbf{The Effects of System Characteristics on Learning}}

Studies have also investigated how search systems and features can impact learning. Studies have considered different system characteristics: (1) the type of device used to search, (2) the presence of novel interface features and tools, and (3) the underlying retrieval algorithm. 

\textbf{Devices:} Demaree et al.~\cite{demaree_influence_2020} compared learning outcomes between participants searching on a smartphone versus laptop computer. Participants were asked to gather information on a controversial topic and write an argumentative essay. Learning was measured by counting the number of pro and con arguments in the essay. While participants issued more queries while searching on a laptop, their learning outcomes were not significantly different across devices.

\textbf{Interface Features and Tools:} Wilson et al.~\cite{wilson_backward_2008} evaluated different interfaces for browsing a music collection. To measure learning, participants enumerated facts learned about the items in the collection. Results found a positive effect on learning from an interface that highlighted item metadata. Kammerer et al.~\cite{kammerer_signpost_2009} evaluated a system that enabled users to use social tags to filter search results. To measure learning, participants completed tests that required them to summarize their knowledge and recall domain-relevant keywords. Participants scored higher on both tests with the experimental system versus a baseline system without social tags. 

Freund et al.~\cite{freund_effects_2016} investigated the impact of two factors on participants' reading comprehension of pre-selected articles: (1) whether articles were displayed in plain text versus HTML, which included distracting elements (e.g., ads), and (2) whether participants could add ``sticky notes'' to articles.  Without the ``sticky notes'' tool, participants had higher reading comprehension scores in the plain text versus HTML condition.  Conversely, with the ``sticky notes'' tool, participants performed equally well in both conditions.

Syed et al.~\cite{syed_improving_2020} evaluated an experimental system that combined eye-tracking and an automatic question-generation feature. The system was designed to ask automatically generated questions about passages read by the searcher during the session. To measure learning, participants completed short-answer tests before each search task, immediately after, and one week later (to measure retention). Participants using the experimental system achieved higher retention scores than participants in the control group.

Qiu et al.~\cite{qiu_towards_2020} investigated the impact of two factors on learning and retention: (1) traditional vs. conversation search and (2) note taking. To measure learning and retention, participants completed the same test before each search task, immediately after, and three days later. Participants achieved greater knowledge gains in the traditional (vs. conversational) search condition, and this effect was stronger when participants could take notes. Interestingly, while knowledge gains were lower in the conversational search condition, participants had greater \emph{retention rates}.

Roy et al.~\cite{roy_note_2021} investigated the impact of two tools that allowed participants to highlight passages and take notes. To measure learning, participants wrote post-task knowledge summaries that were analyzed based on the number of facts included and subtopics covered. Results found benefits from each tool in isolation. Specifically, the note-taking tool enabled participants to write summaries with more facts and the highlighting tool enabled participants to write summaries covering more subtopics. Interestingly, participants did not exhibit greater knowledge gains when using \emph{both} tools, possibly due to cognitive overload.

C\^{a}mara et al.~\cite{camara_searching_2021} evaluated different interface features to support learning: (1) displaying subtopics in the task domain and (2) displaying the searcher's level of coverage across subtopics during the session. Interestingly, these novel features did not significantly improve learning. Instead, they influenced participants to explore more subtopics \emph{superficially}. As evidence, when given feedback about their topical coverage, participants viewed more search results but had shorter dwell times. Importantly, this trend suggests that feedback features can have unintended effects---they can influence searchers to pursue strategies that undermine their \emph{depth} of learning.

To summarize, prior work has investigated a wide range of tools to improve learning during search. In general, results have found benefits from interfaces that: (1) convey more information about the items in the collection~\cite{kammerer_signpost_2009, wilson_backward_2008}, (2) enable searchers to annotate documents~\cite{qiu_towards_2020,roy_note_2021,freund_effects_2016}, and (3) enable searchers to self-assess their understanding of material read during the session~\cite{syed_improving_2020}. On the other hand, results also suggest that tools can have unintended effects.  For example, they can lead to cognitive overload~\cite{roy_note_2021} and encourage searchers to cover more information \emph{superficially}~\cite{camara_searching_2021}.

\textbf{Retrieval Algorithms:} Early work by Hersh et al.~\cite{hersh_towards_1995} evaluated two search systems (i.e., Boolean vs.~TF.IDF retrieval) based on their ability to help medical students improve their performance on a short-answer test. Both systems performed equally well, suggesting that people can achieve comparable learning outcomes using systems that afford very different search strategies. In the context of vocabulary learning, Syed and Collins-Thompson~\cite{syed_optimizing_2017} evaluated a retrieval algorithm that favored documents with a greater \emph{density} of target vocabulary words. Participants had better learning outcomes with the experimental versus baseline system. Weingart and Eickhoff~\cite{weingart_retrieval_2016} explored the impact of several well-established retrieval techniques on learning. To measure learning, participants completed multiple-choice tests after each task. Query expansion had a negative effect on learning, possibly due to topic drift from the original query. On the other hand, passage (vs. document) retrieval had a positive effect on learning, possibly because passages have a higher \emph{density} of query-related content than whole documents.

\vspace{-.2cm}
\subsubsection{\textbf{The Relation between Search Behaviors and Learning}} 
\vspace{-.1cm}

Finally, studies have investigated how specific search behaviors relate to learning outcomes. Several studies have investigated how learning outcomes are related to behaviors that could be potentially captured by a search system
~\cite{gadiraju_analyzing_2018, collins-thompson_assessing_2016,yu_predicting_2018,liu_how_2018,lu_personalized_2017,abualsaud_learning_2017,bhattacharya_measuring_2019,palani_active_2021}. Studies have found that searchers with better learning outcomes have a tendency to: (1) spend more time reading documents~\cite{collins-thompson_assessing_2016,gadiraju_analyzing_2018,yu_predicting_2018, lu_personalized_2017}; (2) issue queries with more advanced and uncommon vocabulary~\cite{collins-thompson_assessing_2016,gadiraju_analyzing_2018,bhattacharya_measuring_2019}; (3) issue more diverse queries within the session~\cite{palani_active_2021}; (4) click on search results with longer titles~\cite{yu_predicting_2018}; (5) review more search results that are relevant~\cite{collins-thompson_assessing_2016} and novel~\cite{abualsaud_learning_2017}; and (6) visit sources that are more suitable to the task, such as encyclopedic sources during \emph{receptive} tasks and Q\&A sources during \emph{critical} tasks~\cite{liu_how_2018}.

Other studies have considered search behaviors that are more difficult to capture within existing search environments but are nonetheless insightful. Using eye-tracking, Bhattacharya et al.~\cite{bhattacharya_measuring_2019} found that participants with better learning outcomes had fewer eye regressions (i.e., less re-reading of text). Lei et al.~\cite{lei_effect_2015} examined the search behaviors of 5th graders in the context of a mock school assignment involving video search. An analysis of post-search interviews found that students with better learning outcomes engaged in more metacognitive planning (e.g., setting goals), monitoring (e.g., tracking progress), and evaluating (e,g., reconsidering strategies).

Prior work has also investigated learning outcomes during collaborative search. Several studies have compared learning outcomes from searchers working individually versus in pairs~\cite{chi_exploring_2016,palani_active_2021}. Chi et al.~\cite{chi_exploring_2016} found that participants working in pairs issued more complex queries (i.e., not issued by other participants), which was interpreted as evidence of learning. Palani et al.~\cite{palani_active_2021} found that participants working in pairs did \emph{not} achieve better outcomes. However, the authors noted that the time limit imposed on search tasks may have prevented pairs from overcoming the overhead of collaboration. Xu et al.~\cite{xu_how_2020} found that collaborators with better learning outcomes engaged in more ``division of labor'' strategies and communicated more frequently.

\vspace{-.2cm}
\subsection{The Effects of Task Characteristics on Behaviors and Outcomes}
\vspace{-.1cm}
In our study, participants completed learning-oriented search tasks with objectives that varied by cognitive process (apply, evaluate, create) and knowledge type (factual, conceptual, procedural). In this respect, our research builds on prior work aimed at understanding how task characteristics can influence searchers.

Search tasks have been studied from many different perspectives. Based on an extensive literature review, Li and Belkin~\cite{ Li2008} proposed a unifying framework for characterizing search tasks in terms of \emph{generic facets} and \emph{common attributes}. Generic facets relate to external factors (e.g., self-imposed vs. assigned). On the other hand, common attributes relate to the task itself, and include subjective attributes (e.g., difficulty) and objective attributes (e.g., complexity). 

Our research builds heavily on prior work on the effects of task complexity, which is defined as an inherent property of the task. Studies have investigated task complexity from different perspectives. Wildemuth et al.~\cite{wildemuth2014untangling} conducted an extensive review of the different ways task complexity has been characterized in prior work.  A few common themes emerged from this review.  Specifically, complex tasks involve: (1) more subtasks and/or subtopics; (2) greater uncertainty about aspects of the task (e.g., inputs and outputs); and (3) more complex mental processes.

Closely related to our research, prior studies have investigated task complexity from the perspective of \emph{cognitive} complexity, which relates to the types of mental processes associated with the task. To this end, studies have leveraged the A\&K taxonomy~\cite{kelly2015development,wu2012grannies,capra_differences_2015,brennan_effect_2014,Hu2017,Thomas2015}. Importantly, however, these studies have leveraged the cognitive process dimension and \emph{ignored} the knowledge type dimension. Results from these studies have found that cognitively complex tasks are perceived to be more difficult~\cite{kelly2015development,wu2012grannies,capra_differences_2015,brennan_effect_2014,Hu2017}, require more search activity~\cite{kelly2015development,wu2012grannies,capra_differences_2015,brennan_effect_2014,Hu2017,Thomas2015}, and lead to more \emph{divergent} strategies by searchers performing the same task~\cite{kelly2015development}. 

\vspace{-.2cm}
\subsection{Contributions of Our Research}
\vspace{-.1cm}
Our research in this paper extends prior work in three important ways. First, as illustrated in our review, most ``search-as-learning'' studies have focused on learning \emph{outcomes}.  In this paper, we focused on the search and learning \emph{process}---the pathways followed by searchers towards an objective.  Understanding the search and learning process provides insights about what searchers do and why.  Additionally, it provides insights about potential search tools to support learning.

Second, several studies have leveraged A\&K's taxonomy to investigate how learning objectives (i.e., the goal of the task) can impact perceptions and behaviors.  Importantly, prior studies have only leveraged the cognitive process dimension and ignored the knowledge type dimension.  In our study, we manipulated objectives by varying both the cognitive process (apply, evaluate, create) and knowledge type (factual, conceptual, procedural) of the objective.

Finally, while A\&K's taxonomy has been used to manipulate and study search tasks, it has not been used to understand the search and learning process.  Therefore, as a methodological contribution, our study shows how the taxonomy can be leveraged to study search sessions from a learning-oriented perspective.

\vspace{-.2cm}
\section{Methods}\label{sec:methods}
\vspace{-.1cm}
To investigate RQ1-RQ3, we conducted a laboratory study with 36 participants (25 female). Participants were recruited using an opt-in mailing list of employees from our university. Participants included 18 student employees and 18 non-student employees, and their ages ranged from 19 to 61 ($M=32.61$, $S.D.=12.82$).

\vspace{-.2cm}
\subsection{Study Overview}\label{subsec:overview}
\begin{figure}[t]
\centering
\hspace*{-.5cm}\includegraphics[width=1.1\columnwidth]{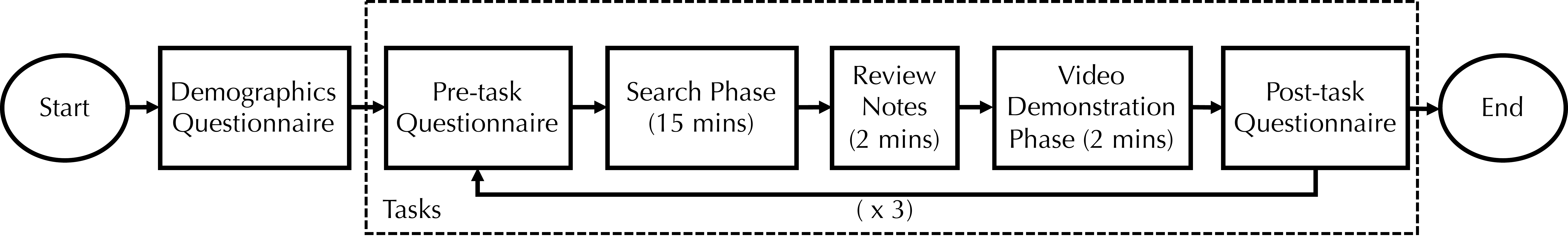}
\vspace{-.5cm}
\caption{\small Study Protocol \vspace{-.5cm}}\label{fig:protocol}
\end{figure}

\vspace{-.1cm}
\textbf{Protocol:} The study protocol is illustrated in Figure~\ref{fig:protocol} and proceeded as follows. After providing informed consent, participants completed a demographics questionnaire. Then, participants completed \emph{three} experimental tasks that followed the same sequence of steps. 

First, after reading the task description, participants completed a pre-task questionnaire. Next, participants completed the \emph{search phase} of the task. During the search phase, participants were given a learning-oriented search task (Section~\ref{subsec:tasks}) and asked to use a custom-built search system to find information and take notes in an external electronic document. The search system was implemented using the Bing Web Search API.\footnote{Given a query, the system returned results from four different verticals in different tabs: web, images, news, and video. Each vertical tab retrieved the top-50 results. Except for the images tab, each tab included pagination controls at the bottom and displayed 10 results per page. The images tab displayed all 50 thumbnails in one SERP using a grid (vs.~list) layout. The Bing API was configured to retrieve results for the US-EN market, and we enabled safe-mode to filter inappropriate results.} In order to investigate participants' pathways towards the learning objective of each task, the study used a think-aloud protocol~\cite{Cooke2010}. Participants' think-aloud comments and search activities were audio/video recorded and later analyzed using qualitative techniques to gain insights about participants' pathways (Section~\ref{subsec:pathways}). Participants were given a maximum of 15 minutes to complete the search phase and were alerted by the moderator when they had 5 minutes remaining. After the search phase, participants were given 2 minutes to review their notes and then completed a 2-minute \emph{video demonstration} phase. During the video demonstration phase, participants were asked to provide a verbal response to the task's main question. Responses were video recorded by the moderator. Finally, after the video demonstration phase, participants were asked to complete a post-task questionnaire. The study session lasted about 1.5 hours and participants were given US\$40 for participating. 

In this paper, we focus on understanding participants' pathways along A\&K's taxonomy towards the learning objective of the task. Thus, responses to the pre-/post-task questionnaires were not analyzed as part of this paper. In Urgo et al.~\cite{UrgoICTIR2020}, we report on the effects of the learning objective on participants pre-/post-task perceptions.  In Section~\ref{sec:discussion}, we discuss how our RQ1-RQ3 results resonate with those reported in Urgo et al.~\cite{UrgoICTIR2020}.

\textbf{Think-aloud:} In order to investigate participants' pathways towards the learning objective of each task, the study used a think-aloud protocol~\cite{Cooke2010}. That is, participants were asked to narrate their thoughts as they searched for information and took notes. The study moderator reminded participants to think aloud if they were silent for an extended period. 

\textbf{Video Demonstration Phase:} During the video demonstration phase of each task, participants were instructed to demonstrate (to whatever extent possible) their achievement of the task's learning objective.  Participants were asked to produce a 2-minute response to the task's main question.  Responses were video recorded by the study moderator. For example, for Task 3 in Section~\ref{subsec:tasks}, participants were asked to verbally demonstrate their novel method for finding the mathematical center of a circle.  Participants' responses were largely verbal. However, participants were also provided with tangible materials as supplemental support: letter-size paper, a large notepad on easel, pens, markers, pencils, eraser, geometric compass, protractor, rulers, and calculator. Participants were informed of (and able to review) these supplemental materials before working on the first experimental task of the study session. The main objective of the video demonstration phase was to encourage participants to learn and discourage them from satisficing.  We believe that asking participants to produce a live demonstration of what they learned achieved this objective.

\vspace{-.2cm}
\subsection{Tasks}\label{subsec:tasks}
\vspace{-.1cm}
Twenty-seven tasks were constructed across three topical domains (art, finance, science). Each domain was associated with nine tasks that varied across three cognitive processes (apply, evaluate, create) and three knowledge types (factual, conceptual, procedural) from A\&K's taxonomy~\cite{anderson2001taxonomy}. To keep the study design manageable, we limited ourselves to three cognitive processes and three knowledge types. 

In terms of cognitive processes, we decided to omit remember, understand, and analyze.  Our goal was to investigate the learning process during complex objectives, which are likely to require pursuing and achieving multiple learning-oriented subgoals. For this reason, we omitted the cognitive processes of remember and understand (i.e., the least complex). Additionally, we decided to omit the cognitive process of analyze for two reasons.  First, we wanted to include objectives that would lead to \emph{divergent} pathways.  To this end, we decided to include evaluate and exclude analyze. Evaluation is closely linked to analysis.  As noted by Anderson and Krathwohl, critiquing elements usually requires first understanding their relations (e.g., similarities and differences)~\cite[p.79]{anderson2001taxonomy}.  Second, Anderson and Krathwohl argue that analyzing is rarely the \emph{ultimate} learning objective in and of itself. Instead, analyzing is ``probably more defensible [as a learning objective] as a prelude to evaluating or creating.''~\cite[p.79]{anderson2001taxonomy}.

In terms of knowledge types, we omitted metacognitive knowledge because it is very different from the other three.  Metacognitive knowledge involves internal (rather than external) knowledge about one's own cognition.\footnote{We also consulted with a cognitive science researcher who advised us to exclude metacognitive knowledge. Understanding metacognitive activities during search-as-learning is an important area for future work.} Next, we provide three example tasks from the science domain: (1) apply/factual, (2) evaluate/conceptual, and (3) create/procedural.

\vspace{-.1cm}
\begin{enumerate}
\item {Apply/Factual---\textbf{Scenario:} You recently watched a TV show about the deepest part of the ocean. The show mentioned the depth of the deepest part of the ocean in meters. However, this number (in meters) did not quite give you clear ``perspective'' on just how deep this is. You want to get a more ``tangible'' appreciation for the depth of the deepest part of the ocean. 

\textbf{Task:} Use the height of the world's tallest building as a unit to measure the deepest part of the ocean.}

\item {Evaluate/Conceptual---\textbf{Scenario:} During a recent trip to the National Air and Space Museum with your extended family, your younger cousin, who is in high school, said she is interested in better understanding how planes are able to fly. You are not very familiar with the principles behind the notion of lift, so when you get home you decide to do some investigating. After some initial research you notice that there are two predominant explanations of lift, Bernoulli's principle and Newton's laws of motion. 

\textbf{Task:} Determine which best explains the notion of lift and why: Bernoulli's principle or Newton's laws of motion? Provide a well-reasoned, logical argument to support your explanation.}

\item {Create/Procedural---\textbf{Scenario:} You are building a firepit in your backyard. You have constructed a large circle so that chairs can fit around the firepit. You have not yet dug the firepit because you want to be sure that it is positioned precisely in the center of the circle. 

\textbf{Task:} Explore different methods for finding the mathematical center of a circle, then create a novel method for finding the mathematical center of your firepit circle. The method can be completely different from those you find, a combination of methods, or a method you find with slight variations.}
\end{enumerate}

As shown, each task was situated in a background scenario that motivated the learning-oriented search task. Task 1 is apply/factual because it requires applying one fact (i.e. the height of the world's tallest building) to gain appreciation of another fact (i.e., the depth of the deepest part of the ocean). Task 2 is evaluate/conceptual because it requires determining which concept (i.e., Bernoulli's principle vs.~Newton's laws of motion) best explains a phenomenon (i.e., lift). Finally, Task 3 is create/procedural because it requires creating a new procedure (i.e., a new method for finding the mathematical center of a circle). Table~\ref{tab:search-tasks} summarizes the main objectives of our 27 tasks. The full tasks are available online: \url{http://www.kelseyurgo.com/tois-2021/}. 

\textbf{Task Assignment:} Each participant completed three tasks from the \emph{same} domain (i.e., arts, finance, or science). Domains were assigned to participants such that 12 participants completed tasks from each domain. Additionally, each participant completed tasks that varied across all three cognitive processes (i.e., apply, evaluate, create) and all three knowledge types (i.e., factual, conceptual, procedural). For example, participant P1 completed the following tasks: (1) apply/factual/art, (2) evaluate/conceptual/art, and (3) create/procedural/art. The ordering of tasks was rotated such that every participant experienced our three cognitive processes and knowledge types in a different order (i.e., 6 CP orders $\times$ 6 KT orders = 36 participants).

\begin{table*}[t]
\captionsetup{justification=centering}
\caption{\small Learning objectives for our 27 tasks. Sub-rows correspond to the domains of science (S), arts (A), and finance (F), respectively. \vspace{-.2cm}}\label{tab:search-tasks}
\centering
\footnotesize
\begin{tabularx}{\textwidth}{|p{1.3cm}|X|X|X|}
\hline
\multirow{2}{\hsize}{Knowledge Dimension} & \multicolumn{3}{c|}{Cognitive Process Dimension} \\
\cline{2-4}
 & Apply & Evaluate & Create\\
 
\hline
\multirow{3}{\hsize}{Factual} & \emph{S:} Use the height of tallest building to measure the deepest part of the ocean. & \emph{S:} Which best describes the deepest part of the ocean: The Mariana Trench or the Challenger deep? & \emph{S:} Generate a table of the ocean's 10 deepest parts including useful and interesting criteria. \\
\cline{2-4}
& \emph{A:} Use the price of the most expensive painting ever sold to express how many Honda civics you could purchase at the same price. & \emph{A:} What is the primary reason for the high price of the most expensive painting ever sold? & \emph{A:} Generate a table of the world's 10 most expensive paintings including their most unique and interesting characteristics. \\ 
\cline{2-4}
& \emph{F:} Use the average retirement savings in the U.S. to calculate how many years one could live on a \$40,000 per year budget. & \emph{F:} Argue whether or not average retirement savings in U.S. is sufficient for determining if citizens are well-prepared for retirement. & \emph{F:} Generate a table of U.S. average retirement savings across demographics you find interesting. \\
\hline
\multirow{3}{\hsize}{Conceptual} & \emph{S:} Use Bernoulli's principle to explain how an airplane flies. & \emph{S:} Which best explains the notion of lift and why: Bernoulli principle or Newton's laws of motion? & \emph{S:} Construct a diagram to illustrate the differences between Bernoulli's principle and Newton's laws of motion applied to lift. \\ 
\cline{2-4}
& \emph{A:} Select which of the following paintings exemplify the ``automatism'' artistic style and note the characteristics indicative of ``automatism''. & \emph{A:} Which artistic movement is more closely related to ``automatism'' and why: Dadaism or Surrealism? & \emph{A:} Construct a graphic that encompasses all artistic forms of ``automatism'' and expresses how each are similar and different. \\
\cline{2-4}
& \emph{F:} Use your knowledge of mortgages to label the following monthly mortgage payment breakdown. & \emph{F:} Which component of a mortgage is most critical for first-time home buyers to negotiate and why: down payment, interest rate, or date of maturity? & \emph{F:} Construct a sample mortgage that would be optimal for your monthly budget and overall savings. \\
\hline
\multirow{3}{\hsize}{Procedural} & \emph{S:} Find the mathematical center of the firepit circle below. & \emph{S:} Find two methods for finding the mathematical center of a circle and choose the best for finding the center of your new firepit and explain why. & \emph{S:} Create a novel method for finding a mathematical center of a firepit circle. \\ 
\cline{2-4}
& \emph{A:} Find a method for making a paper airplane and make that paper airplane. & \emph{A:} Choose which method of paper airplane making would be most fun and interesting and explain why. & \emph{A:} Create a novel method for making a paper airplane. \\ 
\cline{2-4}
& \emph{F:} Use the 50/30/20 budget to categorize the following sample of monthly expenses. & \emph{F:} Explain which budget would be most effective for saving money for your trip and explain why. & \emph{F:} Create a novel budgeting method that is most useful to you in saving money for your vacation.\\
\hline
\end{tabularx}
\vspace{-.7cm}
\end{table*}

\vspace{-.2cm}
\subsection{Qualitative Coding of Pathways}\label{subsec:pathways}
\vspace{-.1cm}
To address RQ1-RQ3, we conducted a qualitative analysis of participants' recorded search sessions, which included their think-aloud comments and search/note-taking activities. The pathway annotation task involved two separate processes.

\textbf{Identifying Learning Instances (LIs):} The first annotation process involved representing each search session as a sequence of \emph{learning instances} (LIs). We define LIs as instances during the search session where the participant either: (1) set forth a new learning-oriented subgoal or (2) serendipitously learned something new and meaningful towards the task's objective. This definition allowed us to account for instances where the participant mentioned pursuing a new specific subgoal (e.g., ``First, I need to find the world's tallest building.''), as well as moments when the participant mentioned serendipitously learning something new without planning (e.g., ``Oh, this says that XYZ was the world's tallest building in 2003. Maybe there's a taller one in 2020.''). To represent each session as a sequence of LIs, one of the authors reviewed all recorded search sessions (i.e., screen activities and think-aloud comments) and marked the timestamp of each LI identified. While recording these timestamps, the question in the author's mind was: At this point, has the searcher set forth a new subgoal or has the searcher learned something new and relevant to the task? Most LIs occurred during points where the participant issued a new query, made a new note in the external document, or engaged with information (e.g., a search result) while thinking aloud. Table~\ref{tab:LI-examples} provides a full example pathway. Each row is a distinct LI. The motivations for marking a new LI are described in the ``Action/Comment'' column.

\textbf{Annotating Learning Instances (LIs):} The second annotation process involved classifying LIs into cells in A\&K's taxonomy---assigning each LI to a specific cognitive process and knowledge type. To classify LIs, we developed a coding guide based on A\&K's book~\cite{anderson2001taxonomy}, which provides definitions and examples of learning objectives for each cognitive process and knowledge type. The full coding guide is available in Appendix~\ref{subsec:coding_guide}. 

The coding guide was developed iteratively using search sessions (i.e., LI sequences) from three participants. First, both of the authors worked together to derive an initial coding guide. Next, both authors independently coded all three sessions from one participant, discussed disagreements, and refined the coding guide. After repeating this process with all three participants, the authors arrived at a final version of the coding guide.

As we developed the coding guide, we found it relatively straightforward to classify LIs into a knowledge type (i.e., factual, conceptual, procedural).  We found it more challenging to classify LIs into a cognitive process (i.e., remember, understand, apply, analyze, evaluate, create).  To alleviate this challenge, the coding guide primarily focuses on criteria for classifying LIs by cognitive process.  Additionally, we found it helpful to use \emph{different wording depending on the LI's knowledge type}.  To illustrate, for classifying LIs as ``understand'', we included the following \emph{analogous} criteria: (1) ``Restates fact in own words.'' (factual); (2) ``Summarizes definition of concept in own words.'' (conceptual); and (3) ``Summarizes steps of procedure in own words.'' (procedural).

To validate the final version of our coding scheme, both authors independently coded all sessions (i.e., LI sequences) from an additional six participants (about 17\% of the data). The Cohen's Kappa agreement was $\kappa=0.809$, which is considered ``almost perfect'' agreement~\cite{Landis77}. Given this high level of agreement, one of the authors (re-)coded all sessions from the remaining 30 participants.

Table~\ref{tab:LI-examples} describes an example annotated pathway, including the action and/or comment that triggered the LI, the A\&K cell associated with the LI, and an explanation/justification of the cell annotation.

\begin{footnotesize}
\begin{longtable}{|p{0.3cm}|p{4.1cm}|c|p{4.5cm}|}
\caption{\small Examples of learning instances (LI) with associated cell assignment and justification. For this pathway, the task's learning objective (evaluate/conceptual) involved deciding which artistic movement is most closely related to Automatism: Dadaism or Surrealism?}\label{tab:LI-examples} \\
 \toprule
 \textbf{LI} & \textbf{Action/Comment} & \textbf{Cell Annotation} & \textbf{Justification} \\
 \midrule
\endfirsthead 
 \toprule
 \textbf{LI} & \textbf{Action/Comment} & \textbf{Cell Annotation} & \textbf{Justification} \\
 \midrule 
\endhead 
 \midrule 
 \multicolumn{4}{r}{\textit{Continued on next page}} \\ 
 \bottomrule
\endfoot 
 \bottomrule
\endlastfoot 
\hline
1 & automatism (query) & Understand/Conceptual & Searcher is gathering information to summarize (understand) automatism (conceptual). \\
\hline
2 & ``The first one that pops up mentions Surrealist automatism, so maybe it has more to do with Surrealism'' & Analyze/Conceptual & Searcher is differentiating and structuring (analyze) automatism (conceptual) within the context of surrealism (conceptual). \\
\hline
3 & ``Maybe we should look at pictures'' & Understand/Conceptual & Searcher is trying to exemplify (understand) automatism (conceptual). \\
\hline
4 & ``It says it was used to express the subconscious...hand is allowed to randomly move across the paper'' & Remember/Conceptual & Searcher is reading a definition (remember) of automatism (conceptual). \\
\hline
5 & subconscious expression (note) & Understand/Conceptual & Searcher is summarizing (understand) a characteristic of automatism (conceptual). \\
\hline
6 & ``Talks about taking material from the subconscious and putting it into art'' & Understand/Conceptual & Searcher is summarizing (understand) a characteristic of automatism (conceptual). \\
\hline
7 & ``Talks about Freud'' & Remember/Factual & Searcher is reading (remember) a fact (factual) related to automatism. \\
\hline
8 & ``Free association'' & Remember/Conceptual & Searcher is reading (remember) about free association (conceptual). \\
\hline
9 & ``This also mentions Surrealism...so let's see what Surrealism and Dadaism are about'' & Analyze/Conceptual & Searcher is differentiating (analyze) between surrealism (conceptual) and dadaism (conceptual). \\
\hline
10 & dadaism art (query) & Understand/Conceptual & Searcher is exploring information to summarize (understand) Dada art (conceptual). \\
\hline
11 & ``An art movement formed during the first world war...a negative reaction to the horrors of the war'' & Remember Conceptual & Searcher is reading an overview (remember) of Dadaism (conceptual). \\
\hline
12 & ``I can compare images of Dadaism'' & Analyze/Conceptual & Searcher is differentiating (analyze) representations of Dadaism from Surrealism (conceptual). \\
\hline
13 & surrealism art (query) & Analyze/Conceptual & Searcher is trying to differentiate (analyze) representations of Surrealist art from Dada art (conceptual). \\
\hline
14 & ``I really want to know how to tell the difference'' & Analyze/Conceptual & Searcher is trying to differentiate (analyze) representations of Surrealist art from Dada art (conceptual). \\
\hline
15 & automatism art (query) & Analyze/Conceptual & Searcher is trying to differentiate (analyze) representations of automatism from Surrealist art and Dada art (conceptual). \\
\hline
16 & ``Automatism doesn't really have like portraits or anything like the other ones'' & Analyze/Conceptual & Searcher is differentiating (analyze) representations of automatism from Surrealist art and Dada art (conceptual). \\
\hline
17 & surrealism art (query) & Understand/Conceptual & Searcher is trying to exemplify (understand) Surrealist art (conceptual). \\
\hline
18 & ``Going to look for more facts about Surrealism'' & Remember Conceptual & Searcher is trying to read (remember) specific information about Surrealism (conceptual). \\
\hline
19 & unconscious mind (note) & Understand/Conceptual & Searcher is gathering information to summarize (understand) Surrealism (conceptual). \\
\hline
20 & ``Has to do with the unconscious mind which is very similar to automatism'' & Analyze/Conceptual & Searcher is differentiating (analyze) between Surrealism and automatism (conceptual). \\
\hline
21 & ``Talks about free association as well'' & Analyze/Conceptual & Searcher is differentiating (analyze) between Surrealism and automatism (conceptual). \\
\hline
22 & ``Not all of this art is abstract'' & Understand/Conceptual & Searcher is identifying (understand) characteristics of Surrealist art (conceptual) \\
\hline
23 & ``Dadaism was reaction to the horrors of the war so maybe Dadaism is less abstract?'' & Analyze/Conceptual & Searcher is differentiating (analyze) Dadaism from Surrealism (conceptual). \\
\hline
24 & ``Have to do with more what they were experiencing consciously rather than subconsciously'' & Analyze/Conceptual & Searcher is differentiating (analyze) Dadaism from Surrealism (conceptual). \\
\hline
25 & ``I think that automatism and Surrealism have more in common'' & Evaluate/Conceptual & Searcher is judging (evaluate) that automatism is more closely related to Surrealism than Dadaism (conceptual). \\
\hline
26 & similarities between automatism and surrealism (query) & Analyze/Conceptual & Searcher is differentiating (analyze) between automatism and Surrealism (conceptual). \\
\hline
27 & processes not under conscious control (note) & Understand/Conceptual & Searcher is summarizing (understand) characteristic of automatism (conceptual). \\
\hline
28 & ``I think that's the biggest similarity'' & Analyze/Conceptual & Searcher is differentiating (analyze) between automatism and Surrealism (conceptual). \\
\hline
29 & interpretation of dreams (note) & Understand/Conceptual & Searcher is summarizing (understand) characteristic of automatism (conceptual). \\
\hline
30 & automatism artists (query) & Understand/Factual & Searcher is exploring (understand) artists (factual) that used automatism. \\
\hline
31 & technique used by surrealist painters (note) & Understand/Conceptual & Searcher is summarizing (understand) characteristic of automatism (conceptual). \\
\hline
32 & ``I don't see how it's not the same thing'' & Analyze/Conceptual & Searcher is differentiating (analyze) between automatism and Surrealism (conceptual). \\
\hline
33 & influenced by freud (note) & Understand/Factual & Searcher is summarizing (understand) isolated unit of information (factual) associated with automatism. \\
\hline
34 & dadaism (query) & Understand/Conceptual & Searcher is gathering information to summarize (understand) Dadaism (conceptual). \\
\hline
35 & ``I don't really know what to describe Dadaism as'' & Understand/Conceptual & Searcher is gathering information to describe (understand) Dadaism (conceptual) in own words. \\
\hline
36 & ``World War I, wasn't an artistic style'' & Understand/Conceptual& Searcher is summarizing (understand) information about Dadaism (conceptual). \\
\hline
\end{longtable}
\end{footnotesize}

\section{Results}\label{sec:results}

In RQ1-RQ3, we analyze the pathways followed by participants towards a specific type of learning objective. A pathway is defined as a sequence of learning instances (LIs) that were each manually assigned to a cell from A\&K's taxonomy (i.e., cognitive process and knowledge type). 

A preliminary analysis of pathways found that participants primarily stayed within the \emph{same} knowledge type as the learning objective's knowledge type. For example, pathways towards factual objectives mostly involved factual LIs. The percentage of LIs associated with the \emph{same} knowledge type as the objective was 97\% for factual objectives, 85\% for conceptual objectives, and 97\% for procedural objectives. Therefore, in our analyses for RQ1-RQ3, we focus \emph{exclusively} on the cognitive processes associated with LIs along the pathway. In other words, while A\&K's taxonomy involves \emph{two} orthogonal dimensions (i.e., cognitive process and knowledge type), pathways were analyzed from a \emph{one}-dimensional perspective (i.e., cognitive process).\footnote{Our analysis of pathways found that participants primarily stayed within the same knowledge type as the given objective.  It is important to emphasize that LIs were classified based on the \emph{type of knowledge} participants were aiming to acquire and \emph{not} solely the \emph{type of information} participants engaged with in pursuit of the subgoal.  To illustrate, imagine an LI in which a searcher is trying to get a basic understanding of Bernoulli's principle.  Now, suppose the searcher encounters and struggles to internalize the following statement: ``Pressure decreases when the speed of a fluid increases.''  This, of course, is a \emph{factual} statement.  However, we would classify this LI as `understand/conceptual' because the current subgoal is to understand the \emph{concept} of Bernoulli's principle.}

Pathways were analyzed from three perspectives: (RQ1) pathway length and number of distinct cognitive processes traversed; (RQ2) types of cognitive processes traversed; and (RQ3) types of transitions between cognitive processes conditioned on the knowledge type of the learning objective. In RQ1 and RQ2, we investigate differences based on the learning objective's cognitive process (apply vs. evaluate. vs. create) and knowledge type (factual vs. conceptual. vs. procedural). To test for significant differences, we used one-way ANOVAs with Bonferroni-corrected post-hoc comparisons.

%Also in RQ1 and RQ2, we used multilevel modeling (MLM) to investigate the effects of learning objective characteristics (knowledge type, cognitive process, and domain), pre-task questionnaire items (interest, prior knowledge, a priori determinability), and post-task questionnaire items (search difficulty and video assessment difficulty) on pathways (pathway length, pathway distinct LIs, and cognitive processes of pathway LIs). All MLMs are grouped by participant to account for random effects that might have been introduced by differences in participants.
\vspace{-.2cm}
\subsection{RQ1: Pathway Characteristics}
\vspace{-.1cm}

In RQ1, we investigate the characteristics of pathways \emph{conditioned} on the learning objective's cognitive process (Table~\ref{tab:pathway-lengths-unique-LOCP}) and knowledge type (Table~\ref{tab:pathway-lengths-unique-LOKT}). Tables~\ref{tab:pathway-lengths-unique-LOCP} \& \ref{tab:pathway-lengths-unique-LOKT} show the average pathway length (i.e., average number of LIs traversed along the pathway) and the average number of \emph{distinct} cognitive processes traversed.

First, we compare pathway characteristics conditioned on the learning objective's cognitive process (Table~\ref{tab:pathway-lengths-unique-LOCP}). The learning objective's cognitive process did not have a significant effect on the pathway length. However, as might be expected, there is an upward trend as the learning objective increases in complexity from apply ($M = 15.97$) to evaluate ($M = 16.89$) to create ($M = 21.17$). In other words, more complex learning objectives had longer pathways (i.e., more learning instances). Similarly, the learning objective's cognitive process did not have a significant effect on the number of distinct cognitive processes traversed along the pathway. However, there is a small upward trend as the complexity of the learning objective increases. Create objectives had pathways with slightly more distinct cognitive processes than apply and evaluate objectives.

Next, we compare pathway characteristics conditioned on the learning objective's knowledge type (Table~\ref{tab:pathway-lengths-unique-LOKT}). The learning objective's knowledge type did not have a significant effect on the pathway length. Conversely, the learning objective's knowledge type did have a significant effect on the number of distinct cognitive processes traversed along the pathway ($F(2,105)=4.75$, $p<.05$). Procedural learning objectives had significantly more distinct cognitive processes than conceptual learning objectives ($p<.01$). We discuss these trends in Section~\ref{sec:discussion}.

\begin{table}[t]
\vspace{-.1cm}
\small
\caption{\small The effects of the learning objective's (LO) \emph{cognitive process} on pathway length and distinct cognitive processes (CPs) traversed (Means $\pm$ 95\% CIs).}
\label{tab:pathway-lengths-unique-LOCP}
\centering
\begin{tabular}{|p{2.9cm}|P{2.0cm}|P{2.0cm}|P{2.0cm}|}
\hline
 & Apply LO & Evaluate LO & Create LO \\ 
\hline
Average Length & $\mathbf{15.97}\pm2.45$ & $\mathbf{16.89}\pm2.52$ & $\mathbf{21.17}\pm4.04$ \\
Average Distinct CPs & $\mathbf{3.72}\pm0.28$ & $\mathbf{3.75}\pm0.26$ & $\mathbf{4.03}\pm0.34$ \\ 
\hline
\end{tabular}
\end{table}

\begin{table}[t]
\small
\caption{\small The effects of the learning objective's (LO) \emph{knowledge type} on pathway length and distinct cognitive processes (CPs) traversed (Means $\pm$ 95\% CIs). $\ddagger$ denotes rows with significant differences across knowledge types: factual (F), conceptual (C), and procedural (P).}
\label{tab:pathway-lengths-unique-LOKT}
\centering
\begin{tabular}{|p{4.0cm}|P{2.0cm}|P{2.0cm}|P{2.0cm}|}
\hline
 & Factual LO & Conceptual LO & Procedural LO \\ 
\hline
Average Length & $\mathbf{19.25}\pm3.97$ & $\mathbf{19.64}\pm2.88$ & $\mathbf{15.14}\pm2.22$ \\
Average Distinct CPs$^\ddagger$ (C < P) & $\mathbf{3.83}\pm0.24$ & $\mathbf{3.53}\pm0.25$ & $\mathbf{4.14}\pm0.35$ \\ 
\hline
\end{tabular}
\end{table}

\subsection{RQ2: Effects of Learning Objectives on Cognitive Processes}

In RQ2, we investigate the effects of the learning objective on the types of cognitive processes traversed along the pathway. For example, are apply LIs more common for some learning objectives than others? Similar to RQ1, we explore differences by conditioning on the learning objective's cognitive process (Table~\ref{tab:proportion-LOCP}) and by conditioning on the learning objective's knowledge type (Table~\ref{tab:proportion-LOKT}). Tables~\ref{tab:proportion-LOCP} \&~\ref{tab:proportion-LOKT} show the average number of LIs (per pathway) associated with each cognitive process. For example, during apply learning objectives (Table~\ref{tab:proportion-LOCP}), pathways had $4.58$ remember LIs on average. During conceptual learning objectives (Table~\ref{tab:proportion-LOKT}), pathways had $9.61$ understand LIs on average.

We start by discussing our results conditioned on the learning objective's cognitive process (Table~\ref{tab:proportion-LOCP}). Our results found five main trends. First, irrespective of the learning objective's cognitive process, remember and understand were the most frequent cognitive processes traversed. Second, the learning objective's cognitive processes had a significant effect on the number of remember LIs traversed ($F(2,105)=5.66$, $p<.005$). Remember LIs were significantly more common during tasks with an objective to create versus apply ($p<.05$) or evaluate ($p<.01$). Third, the learning objective's cognitive process had a significant effect on the number of apply LIs traversed ($F(2,105)=9.47$, $p<.001$). Apply LIs were significantly more common during tasks with an objective to apply versus evaluate ($p<.001$) or create ($p<.005$). Fourth, the learning objective's cognitive process had a significant effect on the number of create LIs traversed ($F(2,105)=4.98$, $p<.01$). Create LIs were significantly more common during tasks with an objective to create versus apply ($p<.01$). Finally, while not statistically significant, analyze and evaluate LIs were more common during tasks with an objective to evaluate versus apply or create.

Next, we discuss our results conditioned on the learning objective's knowledge type (Table~\ref{tab:proportion-LOKT}). Our results found five main trends. First, irrespective of the learning objective's knowledge type, remember and understand were among the most frequent cognitive processes traversed. Second, the learning objective's knowledge type had a significant effect on the number of remember LIs traversed ($F(2,105)=13.08$, $p<.001$). Remember LIs were significantly more common during tasks with an objective involving factual versus conceptual ($p<.01$) or procedural ($p<.001$) knowledge. Third, the learning objective's knowledge type had a significant effect on the number of understand LIs traversed ($F(2,105)=11.69$, $p<.001$). Understand LIs were significantly more common during tasks with an objective involving conceptual versus factual ($p<.001$) or procedural ($p<.005$) knowledge. Fourth, the learning objective's knowledge type had a significant effect on the number of evaluate LIs traversed ($F(2,105)=9.32$, $p<.001$). Evaluate LIs were significantly \emph{less} common during tasks with an objective involving conceptual versus factual ($p<.005$) or procedural ($p<.001$) knowledge. Fifth, the learning objective's knowledge type had a significant effect on the number of create LIs traversed ($F(2,105)=9.10$, $p<.001$). Create LIs were significantly more common during tasks with an objective involving procedural versus factual ($p<.001$) or conceptual ($p<.005$) knowledge.

\begin{table}[t]
\vspace{-.1cm}
\small
\caption{\small The effects of the learning objective's (LO) \emph{cognitive process} on the number of LIs (per pathway) associated with each cognitive process (Means $\pm$ 95\% CIs). $\ddagger$ denotes rows with significant differences across cognitive processes: apply (A), evaluate (E), and create (C).}
\label{tab:proportion-LOCP}
\centering
\begin{tabular}{|l|P{1.6cm}|P{1.6cm}|P{1.6cm}|}
\hline
CP & Apply LO & Evaluate LO & Create LO \\
\hline
Remember$^\ddagger$ (A,E < C)& $\mathbf{4.58}\pm1.28$ & $\mathbf{4.17}\pm1.26$ & $\mathbf{8.33}\pm2.88$ \\ 
Understand & $\mathbf{6.19}\pm1.36$ & $\mathbf{6.19}\pm1.76$ & $\mathbf{6.83}\pm2.35$ \\ 
Apply$^\ddagger$ (A > E,C)& $\mathbf{2.03}\pm0.89$ & $\mathbf{0.47}\pm0.37$ & $\mathbf{0.53}\pm0.28$ \\ 
Analyze & $\mathbf{1.64}\pm0.76$ & $\mathbf{3.14}\pm1.16$ & $\mathbf{2.78}\pm0.84$ \\ 
Evaluate & $\mathbf{1.58}\pm0.77$ & $\mathbf{2.67}\pm0.55$ & $\mathbf{1.92}\pm0.83$ \\ 
Create$^\ddagger$ (A < C) & $\mathbf{0.08}\pm0.13$ & $\mathbf{0.25}\pm0.32$ & $\mathbf{0.78}\pm0.46$ \\ 
\hline
\end{tabular}
\vspace{-.1cm}
\end{table}

\begin{table}[t]
\small
\caption{\small The effects of the learning objective's (LO) \emph{knowledge type} on the number of LIs (per pathway) associated with each cognitive process (Means $\pm$ 95\% CIs). $\ddagger$ denotes rows with significant differences across knowledge types: factual (F), conceptual (C), and procedural (P).}
\label{tab:proportion-LOKT}
\centering
\begin{tabular}{|l|P{1.8cm}|P{2.0cm}|P{2.0cm}|}
\hline
CP & Factual LO & Conceptual LO & Procedural LO \\
\hline
Remember$^\ddagger$ (F > C,P)& $\mathbf{9.22}\pm2.63$ & $\mathbf{5.13}\pm1.56$ & $\mathbf{2.72}\pm0.93$ \\ 
Understand$^\ddagger$ (C > F,P)& $\mathbf{4.03}\pm2.15$ & $\mathbf{9.61}\pm1.81$ & $\mathbf{5.44}\pm1.01$ \\ 
Apply & $\mathbf{0.89}\pm0.56$ & $\mathbf{1.39}\pm0.81$ & $\mathbf{0.86}\pm0.45$ \\ 
Analyze & $\mathbf{2.58}\pm0.95$ & $\mathbf{2.50}\pm1.05$ & $\mathbf{2.36}\pm0.88$ \\ 
Evaluate$^\ddagger$ (C < F,P)& $\mathbf{2.44}\pm0.82$ & $\mathbf{0.89}\pm0.45$ & $\mathbf{2.83}\pm0.74$ \\ 
Create$^\ddagger$ (F,C < P)& $\mathbf{0.08}\pm0.10$ & $\mathbf{0.11}\pm0.18$ & $\mathbf{0.92}\pm0.51$ \\ 
\hline
\end{tabular}
\end{table}

\subsection{RQ3: Transitions Between Cognitive Processes}
Participants completed three search tasks, each with a learning objective associated with a specific knowledge type---factual, conceptual, or procedural knowledge. In RQ3, we investigate cognitive process transitions \emph{conditioned} on the knowledge type of the learning objective. We discuss our RQ3 results from two perspectives. First, we discuss common trends irrespective of the objective's knowledge type. Second, we discuss trends that are \emph{unique} to learning objectives involving a specific knowledge type (factual vs. conceptual vs. procedural knowledge).

\begin{table}
\small
\caption{\small Each Markov matrix shows the transition probabilities between cognitive processes in learning pathways. Transition probabilities marked with * are \emph{common} across learning objectives. Transition probabilities in bold are \emph{unique} to the particular learning objective shown in that sub-table (each sub-table corresponds to a particular objective knowledge type).}
\begin{subtable}[t]{1\textwidth}
\centering
\begin{tabular}{|l|P{0.8cm}P{0.8cm}P{0.8cm}P{0.8cm}P{0.8cm}P{0.8cm}P{0.8cm}|P{0.8cm}|}
 \hline
 CP & Rem. & Und. & Apply & Ana. & Eval. & Create & End & Count \\ 
 \hline
Start & \textbf{0.47} & \multicolumn{1}{r}{0.44*} & 0.00 & 0.06 & 0.03 & 0.00 & 0.00 & 36 \\
\arrayrulecolor{mygrey}\cline{2-7}
Rem. & \multicolumn{1}{r}{\cellcolor{LightGrey}0.63*} & \textbf{0.11} & 0.03 & 0.13 & 0.08 & \multicolumn{1}{P{0.8cm}|}{0.00} & 0.02 & 332 \\
Und. & \textbf{0.30} & \multicolumn{1}{r}{\cellcolor{LightGrey}0.35*} & 0.03 & \multicolumn{1}{r}{0.15*} & 0.12 & \multicolumn{1}{P{0.8cm}|}{0.01} & 0.04 & 145 \\
Apply & \textbf{0.25} & 0.00 & \multicolumn{1}{r}{\cellcolor{LightGrey}0.39*} & 0.07 & 0.11 & \multicolumn{1}{P{0.8cm}|}{0.00} & \textbf{0.18} & 28 \\
Ana. & \textbf{0.37} & \multicolumn{1}{r}{0.19*} & 0.01 & \multicolumn{1}{r}{\cellcolor{LightGrey}0.21*} & 0.14 & \multicolumn{1}{P{0.8cm}|}{0.01} & 0.07 & 97 \\
Eval. & \textbf{0.20} & \multicolumn{1}{r}{0.26*} & 0.03 & 0.10 & \multicolumn{1}{r}{\cellcolor{LightGrey}0.27*} & \multicolumn{1}{P{0.8cm}|}{0.00} & 0.13 & 88 \\
Create & \textbf{0.33} & \multicolumn{1}{r}{0.33*} & 0.00 & 0.00 & 0.00 & \multicolumn{1}{P{0.8cm}|}{\cellcolor{LightGrey}0.00} & 0.33 & 3 \\
\arrayrulecolor{black}\hline
\end{tabular}
\vspace{0.2cm}
\caption{\small Transition probabilities between cognitive processes across \emph{factual} pathways (i.e., conditioned on the knowledge type of the learning objective).}
\label{tab:CP-transitions-factual}
\end{subtable}

\begin{subtable}[t]{1\textwidth}
\centering
\begin{tabular}{|l|P{0.8cm}P{0.8cm}P{0.8cm}P{0.8cm}P{0.8cm}P{0.8cm}P{0.8cm}|P{0.8cm}|}
 \hline
 CP & Rem. & Und. & Apply & Ana. & Eval. & Create & End & Count \\ 
 \hline
Start & 0.06 & \multicolumn{1}{r}{0.86*} & 0.03 & 0.06 & 0.00 & 0.00 & 0.00 & 36 \\
\arrayrulecolor{mygrey}\cline{2-7}
Rem. & \multicolumn{1}{r}{\cellcolor{LightGrey}0.38*} & 0.44 & 0.06 & 0.05 & 0.02 & \multicolumn{1}{P{0.8cm}|}{0.01} & 0.04 & 185 \\
Und. & 0.25 & \multicolumn{1}{r}{\cellcolor{LightGrey}0.46*} & 0.07 & \multicolumn{1}{r}{0.14*} & 0.04 & \multicolumn{1}{P{0.8cm}|}{0.01} & 0.04 & 346 \\
Apply & 0.18 & \multicolumn{1}{r}{0.50*} & \multicolumn{1}{r}{\cellcolor{LightGrey}0.22*} & 0.04 & 0.02 & \multicolumn{1}{P{0.8cm}|}{0.00} & 0.04 & 50 \\
Ana. & 0.17 & \multicolumn{1}{r}{0.43*} & 0.03 & \multicolumn{1}{r}{\cellcolor{LightGrey}0.21*} & \textbf{0.08} & \multicolumn{1}{P{0.8cm}|}{0.00} & 0.08 & 90 \\
Eval. & 0.06 & \multicolumn{1}{r}{0.22*} & 0.03 & \textbf{0.34} & \multicolumn{1}{r}{\cellcolor{LightGrey}0.19*} & \multicolumn{1}{P{0.8cm}|}{0.00} & \textbf{0.16} & 32 \\
Create & 0.00 & \multicolumn{1}{r}{0.75*} & 0.00 & 0.00 & 0.00 & \multicolumn{1}{P{0.8cm}|}{\cellcolor{LightGrey}0.00} & 0.25 & 4 \\
\arrayrulecolor{black}\hline
\end{tabular}
\vspace{0.2cm}
\caption{\small Transition probabilities between cognitive processes across \emph{conceptual} pathways (i.e., conditioned on the knowledge type of the learning objective).\vspace{-.1cm}}
\label{tab:CP-transitions-conceptual}
\end{subtable}

\begin{subtable}[t]{1\textwidth}
\centering
\begin{tabular}{|l|P{0.8cm}P{0.8cm}P{0.8cm}P{0.8cm}P{0.8cm}P{0.8cm}P{0.8cm}|P{0.8cm}|}
 \hline
 CP & Rem. & Und. & Apply & Ana. & Eval. & Create & End & Count \\ 
 \hline
Start & 0.00 & \multicolumn{1}{r}{0.97*} & 0.00 & 0.00 & 0.03 & 0.00 & 0.00 & 36 \\
\arrayrulecolor{mygrey}\cline{2-7}
Rem. & \multicolumn{1}{r}{\cellcolor{LightGrey}0.22*} & 0.31 & 0.07 & \textbf{0.15} & \textbf{0.15} & \multicolumn{1}{P{0.8cm}|}{0.05} & 0.04 & 98 \\
Und. & 0.20 & \multicolumn{1}{r}{\cellcolor{LightGrey}0.31*} & 0.05 & \multicolumn{1}{r}{0.17*} & \textbf{0.19} & \multicolumn{1}{P{0.8cm}|}{0.05} & 0.03 & 196 \\
Apply & 0.13 & \multicolumn{1}{r}{0.26*} & \multicolumn{1}{r}{\cellcolor{LightGrey}0.19*} & 0.03 & \textbf{0.19} & \multicolumn{1}{P{0.8cm}|}{0.06} & 0.13 & 31 \\
Ana. & 0.15 & \multicolumn{1}{r}{0.22*} & 0.00 & \multicolumn{1}{r}{\cellcolor{LightGrey}0.25*} & 0.24 & \multicolumn{1}{P{0.8cm}|}{0.05} & 0.09 & 85 \\
Eval. & 0.15 & \multicolumn{1}{r}{0.32*} & 0.06 & 0.13 & \multicolumn{1}{r}{\cellcolor{LightGrey}0.21*} & \multicolumn{1}{P{0.8cm}|}{0.06} & 0.08 & 102 \\
Create & 0.12 & \multicolumn{1}{r}{0.33*} & 0.06 & 0.06 & 0.03 & \multicolumn{1}{P{0.8cm}|}{\boldGrey{0.18}} & \textbf{0.21} & 33 \\
\arrayrulecolor{black}\hline
\end{tabular}
\vspace{0.2cm}
\caption{\small Transition probabilities between cognitive processes across \emph{procedural} pathways (i.e., conditioned on the knowledge type of the learning objective).}
\label{tab:CP-transitions-procedural}
\end{subtable}
\end{table}

Tables~\ref{tab:CP-transitions-factual}-\ref{tab:CP-transitions-procedural} show the transition probabilities between cognitive processes along pathways towards a learning objective involving factual knowledge (Table~\ref{tab:CP-transitions-factual}), conceptual knowledge (Table~\ref{tab:CP-transitions-conceptual}), and procedural knowledge (Table~\ref{tab:CP-transitions-procedural}). Tables~\ref{tab:CP-transitions-factual}-\ref{tab:CP-transitions-procedural} are Markov matrices. The values along each row correspond to the probabilities of transitioning from one type of LI to another and therefore sum to one. A ``start'' row and ``end'' column have been added as additional states. The values along the start row correspond to the probabilities of starting the pathway with a specific cognitive process. The values in the end column correspond to the probabilities of ending the pathway with a specific cognitive process. The last column shows the raw counts for each state across pathways conditioned on the knowledge type of the objective. Cells along the diagonal (shown in grey) are transitions to the same cognitive process. Cells above the diagonal, and underneath the grey line, are transitions to a \emph{more} complex cognitive process. We refer to these as transition ``upshifts''. Cells below the diagonal are transitions to a \emph{less} complex cognitive process. We refer to these as transition ``downshifts''.

In our discussion of trends, we focus on high versus low transition probabilities. Thus, an important question is: What is a high transition probability? In other words, what is a logical threshold to distinguish between high and low transition probabilities? As shown in Tables~\ref{tab:CP-transitions-factual}-\ref{tab:CP-transitions-procedural}, each state can transition to seven states (i.e., 6 cognitive process states + the end state). If all transitions were equally likely, then all transition probabilities would be ~0.14 (i.e., 1/7). Therefore, we consider transition probabilities $\geq 0.14$ as ``high'' transition probabilities and transition probabilities $< 0.14$ as ``low'' transition probabilities.

\textbf{Common Trends:} First, we discuss common trends irrespective of the objective's knowledge type. In Tables~\ref{tab:CP-transitions-factual}-\ref{tab:CP-transitions-procedural}, common trends are marked with an asterisk (*). Our results found four common trends. First, understand is a common starting point. All ``Start'' rows have high transition probabilities to understand. Second, downshifts to understand (under the diagonal in the understand column) are generally common. As shown in Figure~\ref{tab:CP-transitions-factual}, the single exception is for learning objectives involving factual knowledge. Factual learning objectives had a low transition probability from apply to understand. Later, we discuss possible explanations for this lower transition probability. Third, most cognitive processes had high transition probabilities back to themselves. These probabilities are shown along the diagonal of each matrix, highlighted in grey. As the exception, create-to-create transitions were only common for procedural learning objectives (i.e., not factual nor conceptual objectives). Finally, transitions from understand to analyze were generally common.%across all knowledge types of learning objectives.

Next, we describe trends unique to factual, conceptual, and procedural learning objectives. In Tables~\ref{tab:CP-transitions-factual}-\ref{tab:CP-transitions-procedural}, unique trends are shown in bold.

\textbf{Trends Unique to Factual Learning Objectives:} Our results found four trends unique to factual learning objectives (Table~\ref{tab:CP-transitions-factual}). First, starting with remember was likely only for factual learning objectives. Second, downshifts to remember from \emph{all} cognitive processes (remember column in Table~\ref{tab:CP-transitions-factual}) were likely only for factual learning objectives. In other words, for conceptual and procedural learning objectives, downshifts to remember were likely from \emph{some} cognitive processes but not others. Conversely,  downshifts to remember were likely from \emph{all} cognitive processes during factual learning objectives. Third, transitions from remember to understand were common for conceptual and procedural learning objectives, but \emph{uncommon} for factual learning objectives. Fourth, ending with apply was likely only for factual learning objectives.

\textbf{Trends Unique to Conceptual Learning Objectives:} Our results found three trends unique to conceptual learning objectives (Table~\ref{tab:CP-transitions-conceptual}). First, transitions from evaluate to analyze were likely only for conceptual learning objectives. Second, transitions from analyze to evaluate were common for factual and procedural learning objectives, but \emph{uncommon} for conceptual learning objectives. Third, ending with evaluate was likely only for conceptual learning objectives.

\textbf{Trends Unique to Procedural Learning Objectives:} Our results found three trends unique to procedural learning objectives (Table~\ref{tab:CP-transitions-procedural}). First, upshifts were much more common for procedural learning objectives. Procedural learning objectives had seven likely upshifts compared to only two for factual and two for conceptual learning objectives. Four of these seven upshifts are unique to procedural learning objectives: (1) remember to analyze, (2) remember to evaluate, (3) understand to evaluate, and (4) apply to evaluate. Second, transitions from create to create were likely only for procedural learning objectives. Finally, ending with create was more common for procedural learning objectives. Factual and conceptual learning objectives had high transition probabilities from create to the end state. However, in both cases, these high transition probabilities are based on a single occurrence of create at the end of the pathway (see `Count' column in Tables~\ref{tab:CP-transitions-factual} and~\ref{tab:CP-transitions-conceptual}). Therefore, we do not consider this trend as ``common'' to all learning objectives, but rather a likely transition specific to procedural learning objectives.

\section{Discussion}\label{sec:discussion}

In this section, we summarize our findings and compare them to findings from prior work.

\subsection{RQ1: Effects on Pathway Length and Diversity of Cognitive Processes Traversed}

In RQ1, we investigate the effects of the learning objective (i.e., cognitive process and knowledge type) on the pathway length and distinct cognitive processes traversed.

\textbf{Effects of the Learning Objective's Cognitive Process:} The learning objective's cognitive process did \emph{not} have a significant effect on the pathway length (number of LIs traversed) nor the number of distinct cognitive processes traversed along the pathway (Table~\ref{tab:pathway-lengths-unique-LOCP}). In Urgo et al.~\cite{UrgoICTIR2020}, we reported on results from the same study.  Specifically, we reported on the effects of the learning objective (i.e., cognitive process and knowledge type) on participants' pre-/post-task perceptions and search behaviors. The objective's cognitive process did not significantly affect any outcome related to participants' perceptions nor behaviors.  Thus, our RQ1 results in the current paper are consistent with those reported in Urgo et al.~\cite{UrgoICTIR2020}.  

A reasonable follow-up question is: Why did the objective's cognitive process not have a significant effect on the pathway length nor diversity? One possible explanation stems from our choice of learning objectives considered in the study. To keep the study design manageable, we considered learning objectives associated with the cognitive processes of apply, evaluate, and create, which have mid-to-high levels of complexity.  We might have observed greater (and significant) differences on the pathway length and diversity had we \emph{also} considered objectives associated with low-complexity processes (i.e., remember and understand).

While the trend was not significant, our RQ1 results found that more cognitively complex objectives had longer (and slightly more diverse) pathways (apply $<$ evaluate $<$ create). In general, this trend resonates with prior studies, which have consistently found that more cognitively complex objectives require more search activity~\cite{kelly2015development,wu2012grannies,capra_differences_2015,brennan_effect_2014,Hu2017,Thomas2015}. Thus, our RQ1 results suggest that some of this increase in search activity may be due to the objective requiring more learning instances along the pathway. 

\textbf{Effects of the Learning Objective's Knowledge Type:} The learning objective's knowledge type had a significant effect on the number of distinct cognitive processes traversed along the pathway (Table~\ref{tab:pathway-lengths-unique-LOKT}). Specifically, procedural objectives had pathways involving a more diverse set of cognitive processes. Based on our RQ2 results (discussed in Section~\ref{subsec:discusson_rq2}), this trend is due to procedural objectives having more create LIs. As shown in the last row in Table~\ref{tab:proportion-LOKT}, procedural objectives had about 1.0 create LIs on average. In contrast, factual and conceptual objectives had about 0.10 create LIs on average. 

In our qualitative coding of pathways, we observed that procedural objectives involved more creative subgoals. Specifically, during procedural objectives, create LIs included instances of the participant: (1) simplifying a procedure by skipping steps; (2) modifying steps to fit the given scenario; (3) changing the implementation of a procedure by using different materials (i.e., those readily available); (4) combining steps from multiple procedures to develop a new procedure; (5) using concepts as inspiration to develop a new procedure; and (6) using innovative techniques to improve the accuracy of a procedure. The presence of more create LIs during procedural objectives is a trend that we also observed in our RQ3 results (discussed in Section~\ref{subsec:discussion_rq3}).

In general, there are several possible reasons for why procedural objectives had more create LIs.  First, procedural knowledge is knowledge about how to perform a specific task.  Tasks can often be accomplished in many different ways.  For example, there are many ways for finding the center of a circle, making a paper airplane, or budgeting expenses.  Therefore, procedural objectives may often provide searchers with the \emph{flexibility} to engage in creative learning processes (e.g., modify procedures). Second, searchers may often have individual preferences that influence them to engage in creative processes.  For example, searchers may be inclined to skip or modify steps based on their individual skills or prior knowledge.  Finally, searchers may have situational constraints that require them to engage in creative processes.  Such constraints may include temporal constraints, available resources, and success criteria (e.g., efficiency vs.~accuracy).

\subsection{RQ2: Effects on Cognitive Processes Traversed}\label{subsec:discusson_rq2} 

In RQ2, we investigate the effects of the learning objective (i.e., cognitive process and knowledge type) on the types of cognitive processes traversed along the pathway.

\textbf{Effects of the Learning Objective's Cognitive Process:} We begin by discussing our results conditioned on the objective's cognitive process (Table~\ref{tab:proportion-LOCP}). Our results found four main trends.

First, irrespective of the objective's cognitive process, remember and understand LIs (i.e., the simplest cognitive processes) were the most common. This indicates that, regardless of the learning objective, remembering and understanding are frequent and important activities that support more complex learning subgoals. For example, before \emph{analyzing} the relations between concepts A and B, it is necessary to first \emph{understand} A and B in isolation. This process may involve several iterations---exploring definitions of A, examples of A, definitions of B, examples of B, etc. Additionally, understanding A and B may involve exploring definitions 
and examples from \emph{different} perspectives.  In terms of definitions, searchers may engage with textual overviews, visual representations, and even mathematical formulas.  In terms of examples, searchers may engage with different types of examples in order to infer common themes.

Second, remember LIs were significantly more common during create learning objectives. We believe this may be an unintended effect of our three create/factual tasks (see Table~\ref{tab:search-tasks}), which asked participants to create a table of facts (i.e., a novel \emph{representation} of factual knowledge). This particular task characteristic may have resulted in more remember LIs during create objectives.

Third, apply LIs were significantly more common during apply learning objectives and, similarly, create LIs were significantly more common during create learning objectives. One possible explanation (supported by Table~\ref{tab:proportion-LOCP}) is that apply and create LIs are \emph{generally uncommon}.  In this respect, it is reasonable that apply and create LIs are \emph{more common} when the ultimate objective is to apply and create, respectively. Apply and create processes are generally not supported by existing search environments. Apply processes include \emph{executing} and \emph{implementing} (e.g., using facts to perform a calculation, using a concept to explain a phenomenon, or using a procedure to perform a task). Apply processes often require a ``tangible'' application of knowledge that is difficult to carry out within the search environment. Similarly, create processes are also tangible in nature, involving ``the construction of an original product''~\cite[p.85]{anderson2001taxonomy}. Create processes require tools that support structuring and synthesis in order to generate something new that can be observed (e.g., creating a new table of facts, a concept map, or a step-by-step procedure).

Finally, analyze and evaluate LIs were most frequent during evaluate learning objectives. It is quite natural for evaluate objectives to involve more evaluate LIs. It is \emph{also} natural for evaluate objectives to involve more analyze LIs. Anderson and Krathwohl support this connection between analyze and evaluate, noting that analysis is often a \emph{prelude} to evaluation~\cite{anderson2001taxonomy}. Analysis precedes evaluation because it provides the \emph{evidence} needed for a logical evaluation. For example, in order to evaluate which concept best explains a phenomenon, it is necessary to first analyze the relationships between the different concepts under consideration. This process involves decomposing, comparing, and contrasting (i.e., analyze-level processes).

\textbf{Effects of the Learning Objective's Knowledge Type:} Next, we discuss our results by conditioning on the learning objective's knowledge type (Table~\ref{tab:proportion-LOKT}), Our results found five main trends. 

First, irrespective of the objective's knowledge type, remember and understand LIs were the most common. As previously discussed, this result suggests that remembering and understanding are important processes regardless of the end goal.

Second, remember LIs were most frequent when the objective involved factual knowledge. Factual knowledge is made up of bits of information that tend to be concrete (vs.~abstract) and objective (vs.~subjective) . Unlike other types of knowledge, facts are often immediately comprehended and do not benefit from multiple rounds of summarization and exemplification (i.e., understand-level processes).  Additionally, facts tend to be self-evident (e.g., ``Currently, the world's tallest building is the Burj Khalifa in Dubai.''). Thus, comprehending a fact may not require multiple rounds of decomposition and comparison (i.e., analyze-level processes). Because of these characteristics, we argue that remember is the most common cognitive process for acquiring factual knowledge. This trend resonates with results reported in Urgo et al.~\cite{UrgoICTIR2020}. During factual learning objectives, participants perceived the task to require \emph{more} memorization and \emph{less} activity across cognitive processes more complex than remember.

Third, understand LIs were most frequent when the objective involved conceptual knowledge. We believe this is due to the inherent nature of conceptual knowledge. Conceptual knowledge is knowledge about concepts, categories, theories, principles, schemas, and models.  In this respect, concepts can be highly abstract (e.g., laws of physics) and even subjective (e.g., artistic movements). The amorphous nature of conceptual knowledge may have required more summarization and exemplification (i.e, understand-level processes) to help delineate its boundaries. Furthermore, in our qualitative coding of pathways, we noticed that a single understand LI was often insufficient for acquiring conceptual knowledge. More often, participants iterated on different understand-level activities (e.g., summarizing a definition, exploring an example, sharpening a definition with new information, etc.). This trend also resonates with results reported in Urgo et al.~\cite{UrgoICTIR2020}. During conceptual learning objectives, participants perceived the task to require more understanding.

Fourth, evaluate LIs were the \emph{least} frequent when the objective involved conceptual knowledge. This trend is consistent with the previous trend---conceptual objectives had the most understand LIs. As shown in Table~\ref{tab:proportion-LOKT}, during conceptual objectives, participants spent more LIs iterating on cognitive processes less complex than evaluate (especially understand).  In other words, our results suggest that in order to engage in evaluate-level processes during conceptual objectives, participants had to spend many LIs engaging with \emph{lower}-level processes.

Finally, create LIs were most frequent when the objective involved procedural knowledge. This may be explained by procedural learning objectives allowing for the opportunity to create. As previously discussed, during procedural objectives, participants tended to explore and then modify procedures in order to fit their personal preferences and constraints (e.g., available tools and materials). This trend also resonates with results reported in Urgo et al.~\cite{UrgoICTIR2020}. During procedural objectives, participants perceived the task to require more creating.

\vspace{-.2cm}
\subsection{RQ3: Transitions Between Cognitive Processes Traversed}\label{subsec:discussion_rq3}
\vspace{-.1cm}
In RQ3, we investigate how participants \emph{transitioned} between cognitive processes in their pathways. Our RQ3 results considered trends from two perspectives.  First, we considered common transitions irrespective of the objective.  Second, we considered trends specific to the objective's knowledge type (i.e., factual vs.~conceptual vs.~procedural). Tables~\ref{tab:all-transitions-diagrams}-\ref{tab:uncommon-transitions-diagrams} summarize the main trends observed in our RQ3 results. 

In the following sections, we provide example pathways that demonstrate high-probability transitions. Our goal is to give the reader a sense of why these transitions happened frequently.  In each example, we provide: (1) an overview of the learning objective; (2) a narrative of the learning instances (LIs) shown in the example; and (3) an explanation for why the transition(s) shown in the example may be common during learning-oriented search tasks.

%%%%%%%%%%%%%%%%%%%%%%%%%%%%%%%%%
% COMMON TO ALL OBJECTIVES
%%%%%%%%%%%%%%%%%%%%%%%%%%%%%%%%%

\newcolumntype{f}{>{\hsize=.65\hsize}X}

\begin{table}[h]
\vspace{-.1cm}
\small
\caption{\small Transitions between cognitive processes that were common across all learning objectives. In the transition diagrams,  R = remember, U = understand, P = apply, A = analyze, E = evaluate, and C = create.\vspace{-.3cm}}
\label{tab:all-transitions-diagrams}
\centering
\begin{tabularx}{425pt}{|f|X|}
\hline
\textbf{Transition Diagram} & \textbf{Explanation} \\ 
\hline
\vspace{.1cm}
\hspace{.1cm}
\raisebox{-\totalheight}{\includegraphics[width=0.23\textwidth, height=11mm]{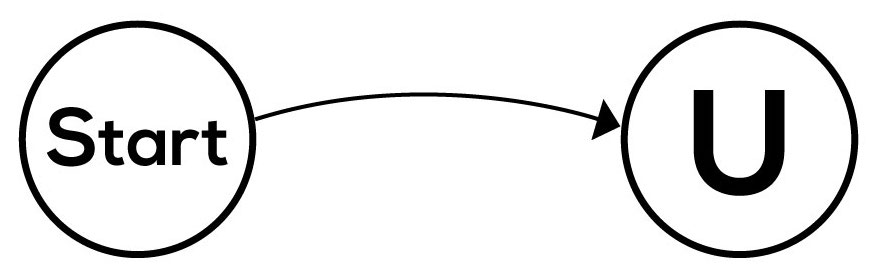}} \newline \centering Starting with Understand
& Learner issues general query for overview of fact, concept, or procedure. \newline E.g., ``expensive painting'', ``automatism'', or ``paper airplane.'' \\ 
\hline
\raisebox{-\totalheight}{\includegraphics[width=0.30\textwidth, height=16mm]{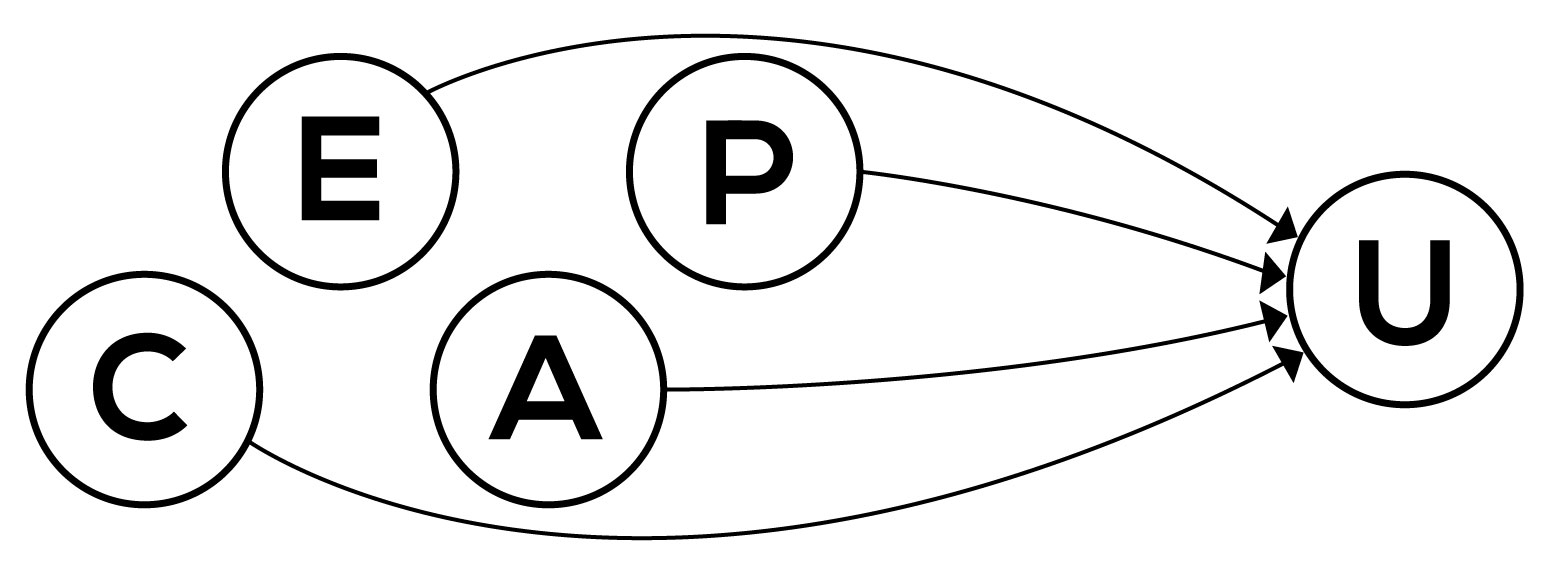}} \newline \centering Downshift to Understand & Learner revisits definition or explores new examples to clarify and deepen understanding. \vspace{.2cm} \newline Additionally, learner gathers more evidence to increase confidence when choosing between options (i.e., evaluate to understand). \\
\hline
\vspace{-.5cm}
\hspace{.3cm}
\raisebox{-\totalheight}{\includegraphics[width=0.3\textwidth, height=20mm]{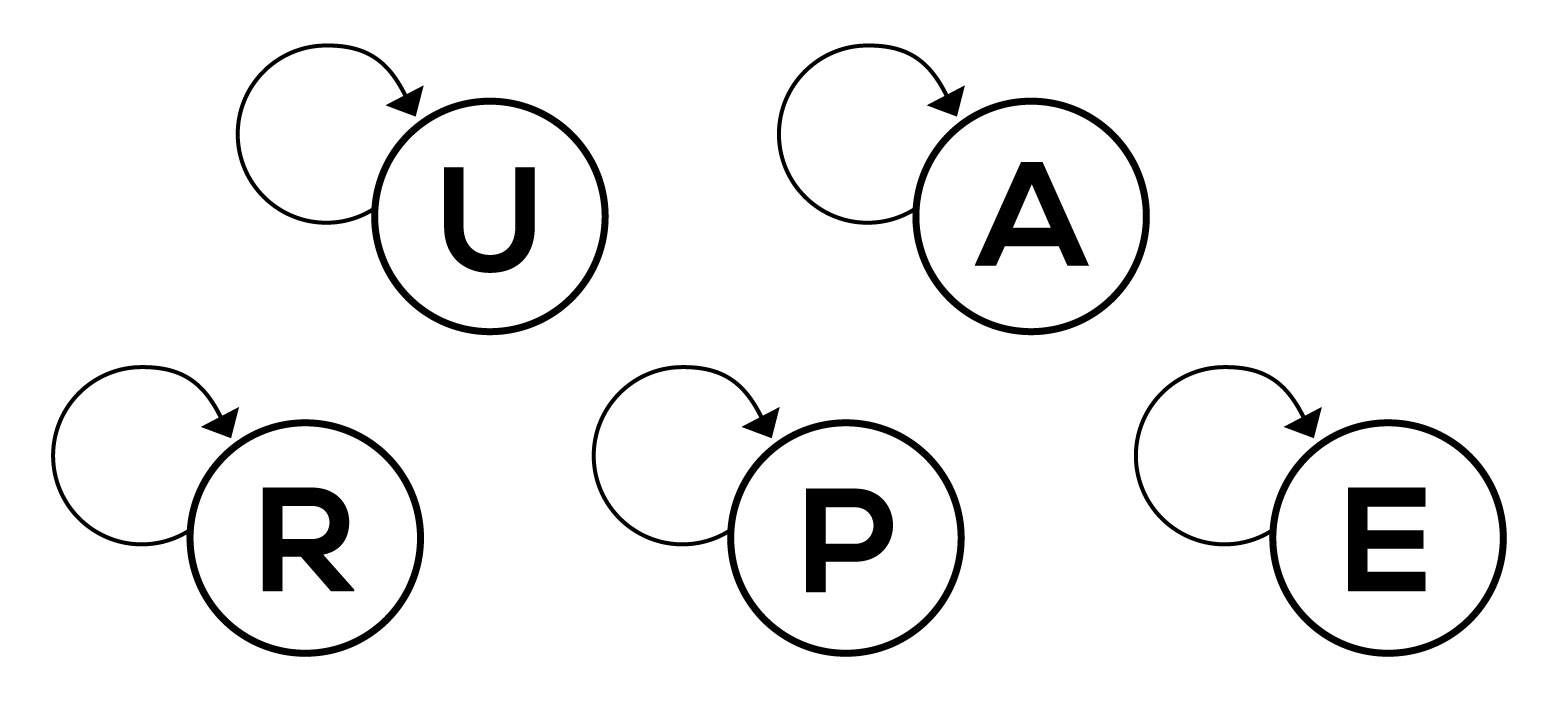}} \newline \centering Back to Self & Learner iterates on the same cognitive process to deepen understanding. \vspace{.2cm} \newline E.g., understand to understand occurs while exploring different characteristics of a concept. \\
\hline
\vspace{.1cm}
\hspace{.1cm}
\raisebox{-\totalheight}{\includegraphics[width=0.24\textwidth, height=12mm]{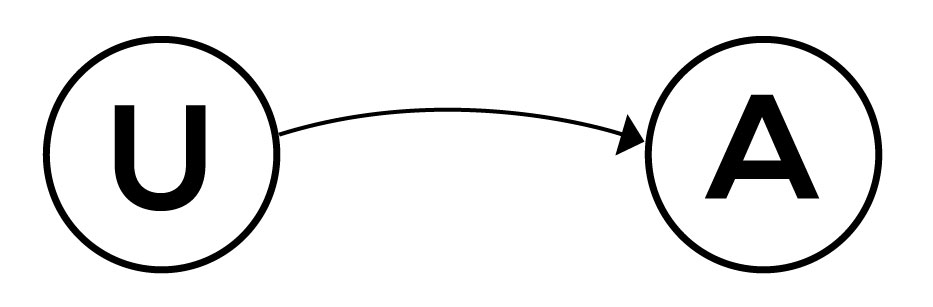}} \newline \centering Understand to Analyze & After acquiring basic or deeper understanding, learner can differentiate from related facts, concepts, and procedures. \\
\hline
\end{tabularx}
\vspace{-.7cm}
\end{table}

\subsubsection{\textbf{High-probability Transitions Across All Learning Objectives}} As illustrated in Table~\ref{tab:all-transitions-diagrams}, our results found four high probability transitions irrespective of the learning objective: (1) starting with understand; (2) downshifting to understand; (3) upshifting from understand to analyze; and (4) transitioning to the same cognitive process. The excerpt following Table~\ref{tab:all-transitions-diagrams} demonstrates all four common LI transitions.

\vspace{.3cm}
\begin{center}
\setlength{\fboxsep}{.5em}
\noindent\fbox{\begin{minipage}{\dimexpr\textwidth-2\fboxsep-2\fboxrule\relax}
    \parbox{\textwidth}{%

\emph{Learning Objective}: The task required the participant to create a diagram of Bernoulli's principle applied to the notion of lift. (create/conceptual)
\begin{itemize}[topsep=8pt]
\footnotesize
 \item \textbf{LI$_1$ (understand)}: Enters query to be able to \underline{summarize} Bernoulli's principle.
 \item ...
 \item \textbf{LI$_{17}$ (understand)}: Reads through \underline{example} diagram of forces involved in Bernoulli's principle applied to a wing during lift.
 \item \textbf{LI$_{18}$ (analyze)}: \underline{Differentiates} those forces from those associated with Newton's Laws applied to lift.
 \item \textbf{LI$_{19}$ (understand)}: Decides they need more information on Newton's laws of motion and next looks for \underline{examples} of newton's laws of motion and lift.
 \item ...
 \item \textbf{LI$_{21}$ (understand)}: Verbally \underline{summarizes} the movement of air above and below a wing in a diagram.
 \item \textbf{LI$_{22}$ (understand)}: \underline{Summarizes} in notes relevant forces at play in the diagram.
 \item ...
 \end{itemize}
     }
\end{minipage}}
\end{center}
\vspace{.3cm}

First, starting with understand (summarizing or exemplifying) is a logical first step in learning. Participants often began the search task with little prior knowledge, leading them to start by issuing a general query with the intent to understand information (e.g., be able to summarize information in their own words).

Second, downshifts to understand are common because summarizing and exemplifying are basic processes that help people construct more precise and nuanced knowledge, which is needed to support more complex processes. In the above example, the participant downshifts from analyze to understand. This downshift occurs because the participant attempts to differentiate between Newton's laws and Bernoulli's principle (LI$_{18}$). In doing so, the participant realizes that they do not comprehend Newton's laws well enough to analyze how they relate to Bernoulli's principle in explaining lift. Downshifting to understand (LI$_{19}$) allowed the participant to review more examples of Newton's laws in order to develop a more nuanced understanding of these concepts.

Third, transitions from understand to analyze are common because better understanding a fact, concept, or procedure enables a learner to analyze how it relates to \emph{other} facts, concepts, or procedures. In the example above, LI$_{17}$ involved clearly defining the boundaries of Bernoulli's principle. Subsequently, LI$_{17}$ enabled LI$_{18}$, which involved analyzing the relationships between Bernoulli's principle and Newton's laws in explaining lift. 

Finally, transitions to the same cognitive process were generally common. The example above shows the particularly ubiquitous understand-to-understand transition. These transitions were common because gaining a deeper understanding of a specific piece of knowledge often involved \emph{iterating} over different definitions, examples, or perspectives. For example, in the above excerpt, LI$_{21}$ shows the participant summarizing the movement of different components in a diagram---the movement of air above and below a airplane's wing during lift. Subsequently, LI$_{22}$ shows the participant making notes about the \emph{forces} acting on the wing during lift (i.e., a different \emph{perspective} on the information depicted in the diagram).

\vspace{-.2cm}
\subsubsection{\textbf{High-probability Transitions Unique to Factual Learning Objectives}} 

As illustrated in Table~\ref{tab:factual-transitions-diagrams}, our results found three high probability transitions unique to factual learning objectives: (1) starting with remember; (2) downshifting to remember; and (3) ending with apply. The excerpt following Table~\ref{tab:factual-transitions-diagrams} demonstrates these three common LI transitions unique to factual objectives.

%%%%%%%%%%%%%%%%%%%%%%%%%%%%%%%%%
% COMMON TO FACTUAL OBJECTIVES
%%%%%%%%%%%%%%%%%%%%%%%%%%%%%%%%%

\begin{table}[h]
\renewcommand\thetable{11}
\vspace{-.1cm}
\small
\caption{\small Transitions between cognitive processes that were common to factual learning objectives. In the transition diagrams,  R = remember, U = understand, P = apply, A = analyze, E = evaluate, and C = create.}
\label{tab:factual-transitions-diagrams}
\centering
\begin{tabularx}{425pt}{|s|X|}
\hline
\centering \textbf{Transition Diagram} & \textbf{Explanation} \\ 
\hline
\vspace{-.5cm}
\hspace{.5cm}
\raisebox{-\totalheight}{\includegraphics[width=0.3\textwidth, height=14mm]{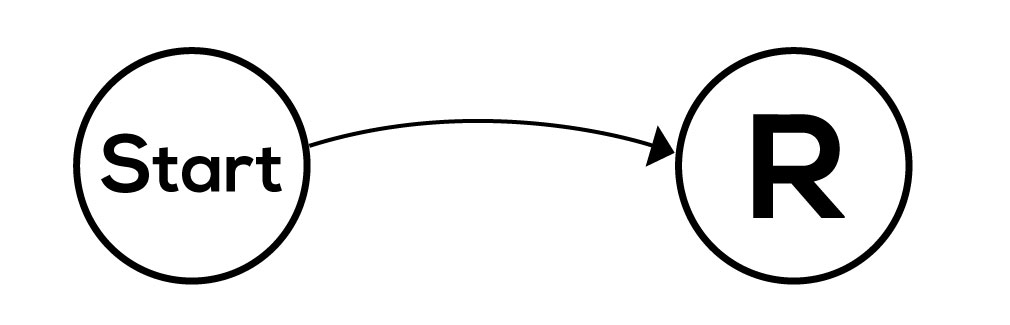}} \newline \centering Starting with Remember
& Learner issues precise, well-defined query of specific fact. \vspace{.3cm} \newline E.g., ``name of most expensive painting'' or ``world's tallest building.'' \\
\hline
\raisebox{-\totalheight}{\includegraphics[width=0.3\textwidth, height=15mm]{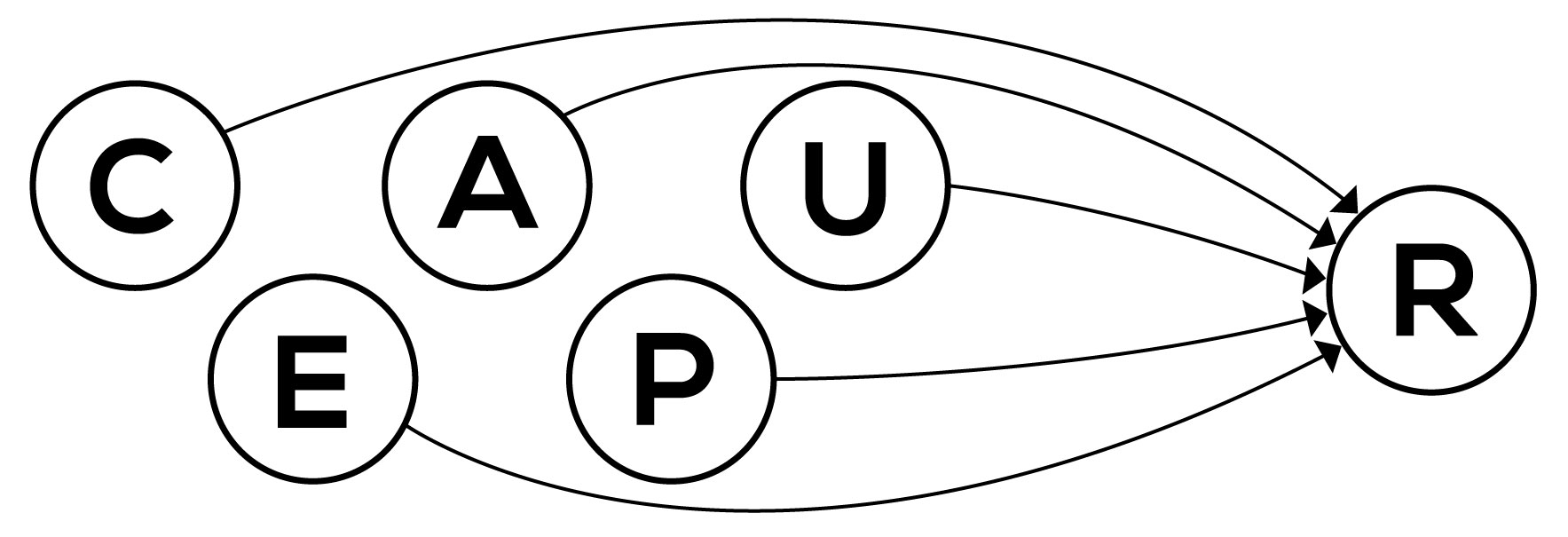}} \newline \centering Downshift to Remember & Learner gathers additional facts. \vspace{.3cm} \newline E.g., learner assesses validity of fact and continues to collect \emph{other} facts. \\
\hline
\vspace{-.5cm}
\hspace{.7cm}
\raisebox{-\totalheight}{\includegraphics[width=0.26\textwidth, height=13mm]{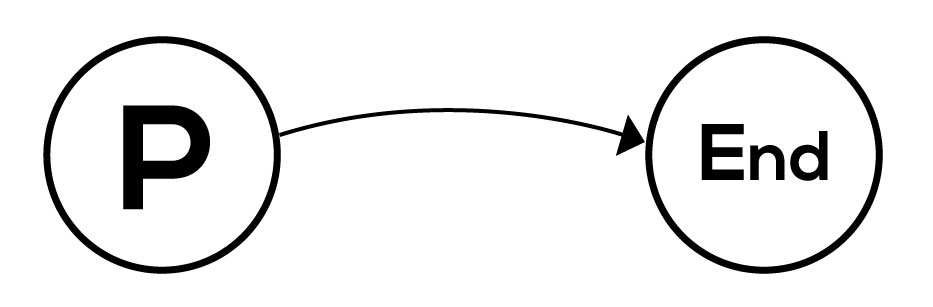}} \newline \centering Apply to End & Learner combines two facts to generate a third fact. \vspace{.3cm} \newline Because facts are well-defined, learners are more likely to be confident in the answer and stop searching. \\
\hline
\end{tabularx}
\vspace{-.3cm}
\end{table}

\vspace{-.4cm}
\begin{center}
\setlength{\fboxsep}{.5em}
\noindent\fbox{\begin{minipage}{\dimexpr\textwidth-2\fboxsep-2\fboxrule\relax}
    \parbox{\textwidth}{%

\emph{Learning Objective}: The task required the participant to use the world's tallest building as a unit to better appreciate the depth of the deepest point in the ocean. (apply/factual)

\begin{itemize}[topsep=8pt]
\footnotesize
 \item \textbf{LI$_1$ (remember)}: Enters a query to \underline{find specifically} the world's tallest building.
 \item ...
 \item \textbf{LI$_5$ (understand)}: \underline{Summarizes} fact in notes about the world's tallest building.
 \item \textbf{LI$_6$ (remember)}: Enters query to be able to \underline{recall} the deepest point in the ocean.
 \item ...
 \item \textbf{LI$_{20}$ (apply)}: \underline{Calculates} the deepest point of the ocean in terms of the world's tallest building.
\end{itemize}
     }
\end{minipage}}
\end{center}
\vspace{0.3cm}

First, remember was a common starting point during factual objectives. Compared to conceptual and procedural knowledge, factual knowledge tends to be concrete, well-defined, and self-contained.  Anderson and Krathwohl noted that factual knowledge relates to bits of information that tend to have ``a low level of abstraction'' and ``value in and of themselves''~\cite[p.42]{anderson2001taxonomy}.  We believe this helps explain why remember was a more common starting point during factual objectives. Participants were often able to gain factual knowledge by simply memorizing or copy/pasting information (versus summarizing or exemplifying). In the above excerpt, in LI$_1$, the participant starts the task by simply searching for the name of the world's tallest building.  

Second, downshifting to remember was a more common transition during factual objectives.  This typically happened when participants sought a new piece of factual knowledge required by the task.  In the above excerpt, the participant downshifts from understand (LI$_5$) to remember (LI$_6$) because they needed a new piece of factual knowledge required by the task (i.e., the name of the deepest point in the ocean).  

Finally, apply was a more common end point during factual objectives.  This common transition is also likely due to factual knowledge being concrete and well-defined.  Applying factual knowledge involves less uncertainty than applying conceptual or procedural knowledge.  In the above excerpt, in LI$_{20}$, the participant is able to apply two bits of factual knowledge to generate a new fact.  Presumably, the participant felt confident enough in their application of this factual knowledge to end the task with apply. This was less common during conceptual or procedural objectives.  When applying conceptual or procedural knowledge, participants often \emph{downshifted} to lower-complexity processes (e.g., revisiting definitions, summaries, and examples) to \emph{verify} whether they applied the knowledge correctly.

\vspace{-.2cm}
\subsubsection{\textbf{High-probability Transitions Unique to Conceptual Learning Objectives}} As illustrated in Table~\ref{tab:conceptual-transitions-diagrams}, our results found two high probability transitions unique to conceptual learning objectives: (1) downshifting from evaluate to analyze and (2) ending with evaluate. The excerpt following Table~\ref{tab:conceptual-transitions-diagrams} demonstrates these two common LI transitions unique to conceptual objectives.

%%%%%%%%%%%%%%%%%%%%%%%%%%%%%%%%%
% COMMON TO CONCEPTUAL OBJECTIVES
%%%%%%%%%%%%%%%%%%%%%%%%%%%%%%%%%

\begin{table}[h]
\renewcommand\thetable{12}
\vspace{-.1cm}
\small
\caption{\small Transitions between cognitive processes that were common to conceptual learning objectives. In the transition diagrams,  R = remember, U = understand, P = apply, A = analyze, E = evaluate, and C = create.}
\label{tab:conceptual-transitions-diagrams}
\centering
\begin{tabularx}{\textwidth}{|Y|X|}
\hline
\textbf{Transition Diagram} & \textbf{Explanation} \\ 
\hline
\hspace{.5cm}
 \raisebox{-\totalheight}{\includegraphics[width=0.3\textwidth, height=15mm]{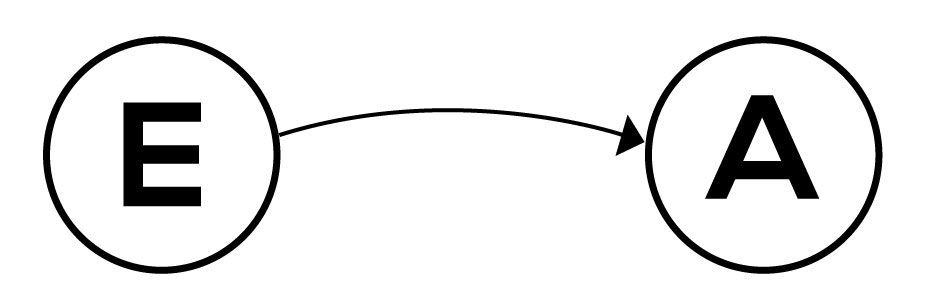}} \newline \centering Evaluate to Analyze
 & Learner gathers more evidence to bolster argument for judgement of concept. \vspace{.3cm} \newline The amorphous, broad nature of concepts often make learners less confident when choosing which concept is most suited to a particular scenario (e.g., explaining a natural phenomenon). \\
 \hline
 \hspace{.5cm}
 \raisebox{-\totalheight}{\includegraphics[width=0.3\textwidth, height=15mm]{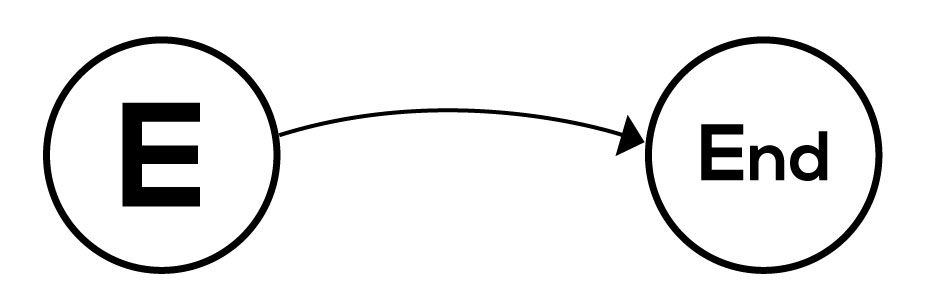}} \newline \centering End with Evaluate & Learners engaged in fewer instances of evaluate overall in conceptual learning objectives. \vspace{.3cm} \newline When it did occur, evaluate was more likely to be the final step after much understanding and analyzing of concepts throughout the pathway. \\
 \hline
\end{tabularx}
\end{table}

\begin{center}
\setlength{\fboxsep}{.5em}
\noindent\fbox{\begin{minipage}{\dimexpr\textwidth-2\fboxsep-2\fboxrule\relax}
    \parbox{\textwidth}{%
    
\emph{Learning Objective}: The task required the participant to determine which art movement most closely aligns with automatism: surrealism or Dadaism. (evaluate/conceptual)

\begin{itemize}[topsep=8pt]
\footnotesize
 \item ...
 \item \textbf{LI$_{32}$ (evaluate)}: Based on characteristics of surrealism and Dadaism, they \underline{judge} that surrealism seems to be more closely aligned with automatism.
 \item \textbf{LI$_{33}$ (analyze)}: In order to further clarify this distinction, they \underline{organize and differentiate} surrealism, Dadaism, and automatism in relation to one another.
 \item ... 
 \item \textbf{LI$_{35}$ (evaluate)}: Using evidence that shows ``surrealist automatism'' as a term, they make a final \underline{judgment} that surrealism is more closely aligned with automatism than Dadaism.
 \end{itemize}
 }
\end{minipage}}
\end{center}

First, downshifting from evaluate to analyze was common during conceptual objectives.  These downshifts mostly occurred when participants felt they needed to reexamine the relations between concepts in order to clarify or strengthen a previously made argument. In the excerpt above, the participant transitions from LI$_{32}$ (evaluate) to LI$_{33}$ (analyze) because they needed to revisit the relations between the three concepts associated with the task in order to make a clearer argument.  

Second, evaluate was a common endpoint during conceptual objectives. This is likely due to the amorphous nature of conceptual knowledge.  In contrast to facts and procedures, concepts (e.g., art movements) can be highly abstract and subjective.  Based on our coding of pathways, we noticed that participants did not feel confident enough to evaluate conceptual knowledge until going through many iterations of understand (e.g., understanding individual concepts) and analyze (e.g., analyzing the relations between concepts).  Our RQ2 results found that evaluate LIs were generally \emph{rare} during conceptual objectives.  Our RQ3 results suggest that participants tended to evaluate towards the \emph{end} of the search session (i.e., the final LI in the pathway).

% \vspace{-.5cm}
\subsubsection{\textbf{High-probability Transitions Unique to Procedural Learning Objectives}} As illustrated in Table~\ref{tab:procedural-transitions-diagrams}, our results found three types of high probability transitions unique to procedural learning objectives: (1) upshifting to more complex processes, (2) transitioning from create to create, and (3) ending with create.

%%%%%%%%%%%%%%%%%%%%%%%%%%%%%%%%%
% COMMON TO PROCEDURAL OBJECTIVES
%%%%%%%%%%%%%%%%%%%%%%%%%%%%%%%%%
\newcolumntype{j}{>{\hsize=.8\hsize}X}
\begin{table}[h]
\renewcommand\thetable{13}
\vspace{-.1cm}
\small
\caption{\small Transitions between cognitive processes that were common to procedural learning objectives. In the transition diagrams,  R = remember, U = understand, P = apply, A = analyze, E = evaluate, and C = create.\vspace{-.3cm}}
\label{tab:procedural-transitions-diagrams}
\centering
\begin{tabularx}{\textwidth}{|Y|X|}
\hline
\textbf{Transition Diagram} & \textbf{Explanation} \\ 
\hline
 \hspace{.4cm}
 \raisebox{-\totalheight}{\includegraphics[width=0.32\textwidth, height=22mm]{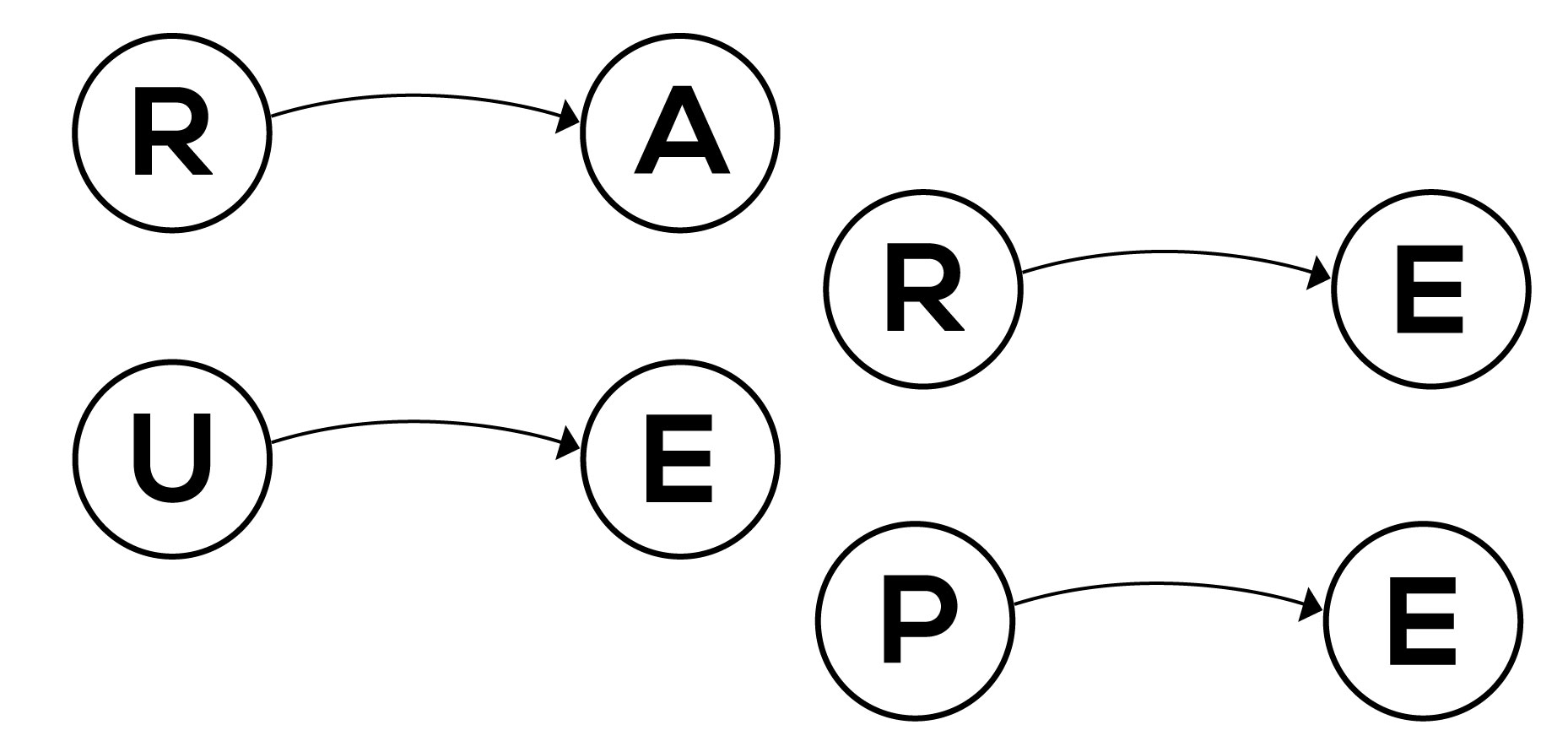}} \newline \centering Upshifts to more complex cognitive processes
 & Learner is able to upshift to analyzing or evaluating without iterating on less complex processes. This may be explained by target procedures being generally well-defined and concrete. \\
 \hline
 \hspace{.6cm}
 \raisebox{-\totalheight}{\includegraphics[width=0.12\textwidth, height=13mm]{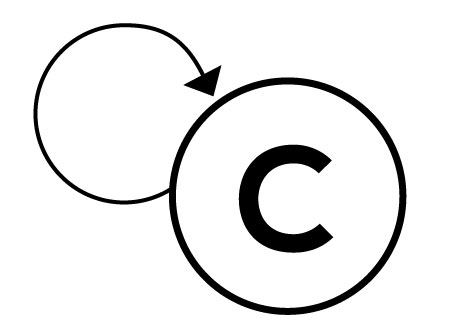}} \newline \centering Back to Create & After creating initial modification, learner is inspired to make additional modifications to procedure (e.g., skip, reorder, modify steps). \\
 \hline
  \hspace{.5cm}
  \raisebox{-\totalheight}{\includegraphics[width=0.24\textwidth, height=12mm]{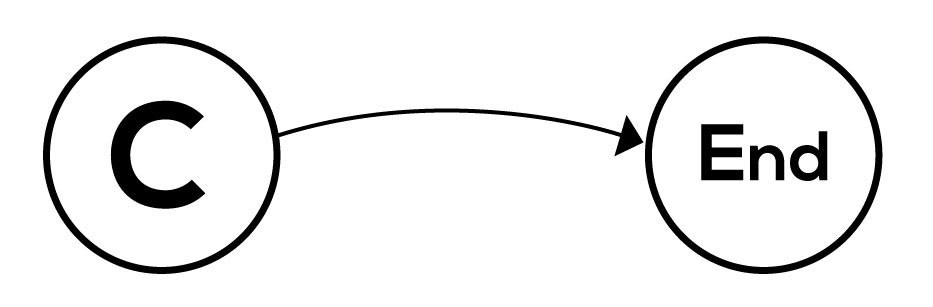}} \newline \centering Create to End & Learner creates novel procedure and ends search to test procedure with tangible materials. \\
 \hline
\end{tabularx}
\vspace{-.1cm}
\end{table}

The following pathway excerpt demonstrates three common transitions unique to procedural objectives: (1) remember to evaluate, (2) understand to evaluate, and (3) remember to analyze. These are the types of transition \emph{upshifts} that were more common during procedural objectives.

\setlength{\fboxsep}{1em}
\fbox{\parbox{\textwidth}{\emph{Learning Objective}: The task required the participant to find the mathematical center of a firepit circle. (apply/procedural)

\begin{itemize}
 \footnotesize
  \item ...
  \item \textbf{LI$_4$ (remember)}: \underline{Reads word-for-word} through a particular method of finding the center of a circle.
  \item \textbf{LI$_5$ (evaluate)}: \underline{Judges} that the method looks too difficult.
  \item \textbf{LI$_6$ (remember)}: \underline{Reads word-for-word} through a different method.
  \item \textbf{LI$_7$ (understand)}: \underline{Summarizes} a list of necessary materials from the steps they just read.
  \item \textbf{LI$_8$ (evaluate)}: \underline{Judges} that this method looks even more complicated.
  \item \textbf{LI$_9$ (remember)}: \underline{Reads word-for-word} through an additional method. 
  \item \textbf{LI$_{10}$ (analyze)}: \underline{Relates} method to previous method, concluding that the methods are similar.
  \item ...
\end{itemize}}}

% \begin{center}
% \setlength{\fboxsep}{.5em}
% \noindent\fbox{\begin{minipage}{\dimexpr\textwidth-2\fboxsep-2\fboxrule\relax}
% \captionof*{}{}
%     \parbox{\textwidth}{%

% \emph{Learning Objective}: The task required the participant to find the mathematical center of a firepit circle. (apply/procedural)

% \begin{itemize}[topsep=8pt]
% \footnotesize
%  \item ...
%  \item \textbf{LI$_4$ (remember)}: \underline{Reads word-for-word} through a particular method of finding the center of a circle.
%  \item \textbf{LI$_5$ (evaluate)}: \underline{Judges} that the method looks too difficult.
%  \item \textbf{LI$_6$ (remember)}: \underline{Reads word-for-word} through a different method.
%  \item \textbf{LI$_7$ (understand)}: \underline{Summarizes} a list of necessary materials from the steps they just read.
%  \item \textbf{LI$_8$ (evaluate)}: \underline{Judges} that this method looks even more complicated.
%  \item \textbf{LI$_9$ (remember)}: \underline{Reads word-for-word} through an additional method. 
%  \item \textbf{LI$_{10}$ (analyze)}: \underline{Relates} method to previous method, concluding that the methods are similar.
%  \item ...
% \end{itemize}
%  }
% \end{minipage}}
% \end{center}

%\pagebreak

First, remember to evaluate was a common upshift during procedural objectives. This suggests that participants were often able to make judgements about procedures using superficial evidence or heuristics (e.g., number of steps, number of tools or materials involved, and complexity of steps depicted visually).  In the excerpt above, the participant reads the steps of a procedure in LI$_4$ and quickly judges that the procedure is too complex in LI$_5$ without trying to fully comprehend it.  

Second, understand to evaluate was a common upshift during procedural objectives. Similar to the previous example, in many cases, participants were able to judge a procedure after understanding some aspect of it.  In the excerpt above, the participant summarizes the materials involved in a procedure in LI$_7$ and subsequently judges that the procedure is too complex in LI$_8$.  

Finally, remember to analyze upshifts were common during procedural objectives.  This trend is likely due to the concrete nature of procedures, especially those involving physical steps and tangible materials. During procedural objectives, participants were often able to draw comparisons between procedures (an analyze process) without first iterating over multiple rounds of understanding.

The next pathway excerpt demonstrates two common transitions unique to procedural learning objectives: (1) create to create and (2) ending with create.

\setlength{\fboxsep}{1em}
\fbox{\parbox{\textwidth}{\emph{Learning Objective}: The task required the participant to find different methods for finding the mathematical center of a circle and then develop their own method. (create/procedural)

\begin{itemize}
\footnotesize
 \item ...
 \item \textbf{LI$_6$ (analyze)}: Explores \underline{alternatives} that can act as a long straightedge in a particular scenario for finding the center of a firepit circle.
 \item \textbf{LI$_7$ (create)}: \underline{Designs} own alternative tool, explaining that they could use string instead of a straightedge and, additionally, use the string as a way to find the shortest distance between two points.
 \item \textbf{LI$_8$ (create)}: \underline{Develops} method of finding the absolute diameter, rather than simply a chord, by initially estimating the diameter then moving a string back and forth to make sure that there isn't a point where the string is longer.
 \item ...
 \item \textbf{LI$_{16}$ (create)}: \underline{Develops} an additional procedure to find the center of the circle using the ``reverse'' of the circumference formula.
\end{itemize}}}

% \begin{center}
% \setlength{\fboxsep}{.5em}
% \noindent\fbox{\begin{minipage}{\dimexpr\textwidth-2\fboxsep-2\fboxrule\relax}
% \captionof*{}{}
%     \parbox{\textwidth}{%

% \emph{Learning Objective}: The task required the participant to find different methods for finding the mathematical center of a circle and then develop their own method. (create/procedural)

% \begin{itemize}[topsep=8pt]
% \footnotesize
%  \item ...
%  \item \textbf{LI$_6$ (analyze)}: Explores \underline{alternatives} that can act as a long straightedge in a particular scenario for finding the center of a firepit circle.
%  \item \textbf{LI$_7$ (create)}: \underline{Designs} own alternative tool, explaining that they could use string instead of a straightedge and, additionally, use the string as a way to find the shortest distance between two points.
%  \item \textbf{LI$_8$ (create)}: \underline{Develops} method of finding the absolute diameter, rather than simply a chord, by initially estimating the diameter then moving a string back and forth to make sure that there isn't a point where the string is longer.
%  \item ...
%  \item \textbf{LI$_{16}$ (create)}: \underline{Develops} an additional procedure to find the center of the circle using the ``reverse'' of the circumference formula.
% \end{itemize}
%  }
% \end{minipage}}

% \end{center}

First, create-to-create transitions were more common during procedural objectives.  There are two possible reasons for this trend.  First, procedures are made up of steps that can be skipped, modified, re-ordered, or combined.  Second, procedures often need to be modified or created according to individual preferences or constraints (e.g., prior knowledge, available tools/materials, and temporal constraints).  As such, procedural objectives may have had more create-to-create transitions because one modification often leads to additional modifications.  In the excerpt above, the participant designs a new tool in LI$_7$ and subsequently modifies a step to use this new tool in LI$_8$.  Finally, create was a common endpoint during procedural objectives.  This trend may simply be due to the prevalence of create LIs during procedural objectives.  After modifying a procedure or creating a new one, it was not necessary for participants to downshift to lower-complexity processes (e.g., understand or analyze) because the new procedure was already fully understood. Thus, participants were more often able to end with create.

\subsubsection{\textbf{Low-probability Transitions Unique to a Knowledge Type}}

Finally, as illustrated in Table~\ref{tab:uncommon-transitions-diagrams}, there were two transitions that were \emph{uncommon} to a particular knowledge type. First, remember-to-understand transitions were uncommon during factual objectives. Most remember transitions simply iterated back to remember.  As noted above, this is likely due to factual knowledge being concrete, well-defined, and self-contained.  We believe these attributes of factual knowledge rendered more complex cognitive processes (e.g., summarizing, exemplifying) unnecessary for comprehension.  Instead, participants tended to iterate from remember to remember LIs as they gathered multiple distinct facts in quick succession.

Second, analyze-to-evaluate upshifts were uncommon during conceptual objectives. Most analyze transitions iterated back to analyze or downshifted to understand.  This trend is likely due to conceptual knowledge being abstract, subjective, and interconnected.  Evaluating conceptual knowledge required more iterations of understand (i.e., understanding individual concepts) and analyze (i.e., analyzing the relations between concepts).  Participants often iterated over these lower-complexity processes before being able to confidently evaluate conceptual knowledge.  This trend resonates with results from Urgo et al.~\cite{UrgoICTIR2020}, which found that participants perceived conceptual objectives to be more difficult than factual and procedural objectives.  Participants' inability to move up in complexity from analyze to evaluate may have contributed to their perceptions of task difficulty.

%%%%%%%%%%%%%%%%%%%%%%%%%%%%%%%%%
% UNCOMMON
%%%%%%%%%%%%%%%%%%%%%%%%%%%%%%%%%

\begin{table}[h]
\renewcommand\thetable{14}
\vspace{-.1cm}
\small
\caption{\small Transitions between cognitive processes that were uncommon. In the transition diagrams,  R = remember, U = understand, P = apply, A = analyze, E = evaluate, and C = create.}
\label{tab:uncommon-transitions-diagrams}
\centering
\begin{tabularx}{\textwidth}{|Y|X|}
\hline
\textbf{Transition Diagram} & \textbf{Explanation} \\ 
\hline
 \hspace{.7cm}
 \raisebox{-\totalheight}{\includegraphics[width=0.25\textwidth, height=12mm]{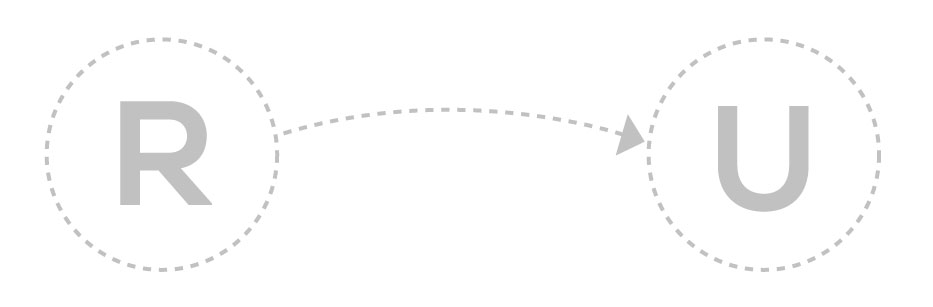}} \newline Remember to Understand (\textit{Factual})
 & Facts often did not require understand (e.g., summarization, exemplification) for learner to internalize \\
 \hline
 \hspace{.7cm}
 \raisebox{-\totalheight}{\includegraphics[width=0.25\textwidth, height=12mm]{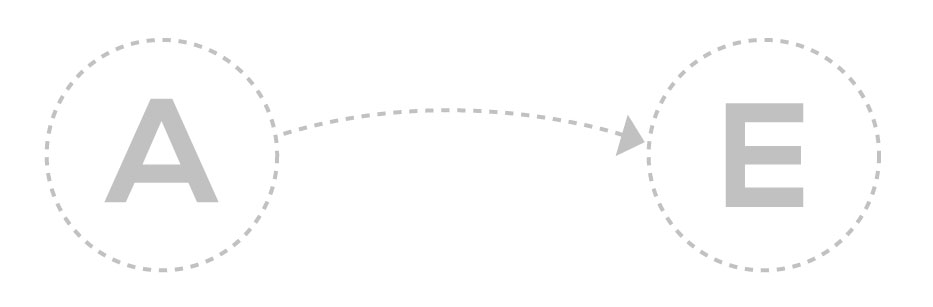}} \newline Analyze to Evaluate (\textit{Conceptual}) & Learners had more difficulty in increasing in complexity when learning concepts \\
 \hline
\end{tabularx}
\end{table}

% \vspace{-0.7cm}
\section{Implications for Designing Tools to Support Learning}

Our results found that the cognitive process, and to a greater extent, the knowledge type of the learning objective impacted: (RQ1) the length and diversity of pathways; (RQ2) the types of cognitive processes traversed along the pathways; and (RQ3) the types of transitions between cognitive processes. Our results have implications for designing search environments that support learning.

In this section, we propose different tools and features to encourage and support learning during search. Table~\ref{tab:implications-summary} provides a summary of our proposed tools. In Table~\ref{tab:implications-summary}, tools are grouped by type of learning objective.  First, we discuss tools to support conceptual objectives.  Then, we discuss tools to support procedural objectives.  Finally, we discuss tools to support \emph{all} objectives. Tools to support factual objectives (e.g., note-taking tools that enable searchers to save information) are also likely to support conceptual and procedural objectives.  Therefore, in Table~\ref{tab:implications-summary}, we did not create a separate section for tools to uniquely support factual objectives. Instead, tools to support factual objectives are discussed under ``All Learning Objectives''. In Table~\ref{tab:implications-summary}, we describe each proposed tool, list which of our results motivate the proposed tool, and acknowledge prior work that has investigated similar tools.

\subsection{Tools to Support Conceptual Learning Objectives}\label{subsec:implications_conceptual}
First, in Table~\ref{tab:implications-summary}, we discuss three potential tools and features to support searchers with conceptual learning objectives.

Our results found that conceptual learning objectives had the most understand LIs (RQ2) and frequent downshifts to understand (RQ3).  As previously noted, during conceptual objectives, participants often felt unsure if they understood concepts well enough to support more complex subgoals (e.g., analyze the relations between concepts or evaluate the relevance of concepts in a given scenario). Participants often iterated over multiple understand LIs (i.e., understand-to-understand transitions) by reviewing definitions, summaries, and examples from different perspectives.  We propose three tools and features to support such understand-level processes during conceptual objectives. 

First, given a conceptual knowledge query (e.g., ``cubism''), systems could diversify and organize search results by definitions, overviews, and examples.  Commercial search systems already do this to some extent (albeit inconsistently).  For example, in response to the query ``cubism'', Google displays dictionary definitions, encyclopedic articles, and images of cubist paintings.  However, systems could organize these different types of results more clearly with explicitly marked sections on the SERP.

Second, to deepen their understanding, participants often explored examples of a concept from different perspectives.  To support these activities, systems could present and organize examples of a concept by \emph{type} within the SERP. To illustrate, Bernoulli's principle (B) can be demonstrated in multiple ways (e.g., B and sail boats, B and wings, B and Venturi tubes, B and chimneys, B and shower curtains, B and curveballs in baseball, and B and topspin in tennis). Grouping examples by \emph{type} may encourage searchers to explore examples from \emph{different} perspectives and learn about \emph{common} themes, deepening their understanding of a concept.

Third, and perhaps more ambitiously, systems could enable searchers to test their own understanding of a concept.  In prior search-as-learning studies, understand-level learning has often been measured using closed-ended tests (e.g., fill-in-the-blank, true/false, and multiple-choice tests).  Such tests could potentially be automated by a search system.  For instance, a system could show a searcher a set of images and ask them to select which ones \emph{exemplify} a given concept (e.g., ``Which of these images exemplify Bernoulli's principle?''). Answering such automatically generated questions may help searchers identify knowledge gaps and assess whether they understand a concept well enough to pursue more complex subgoals (e.g., analyze or evaluate).  Prior research has investigated systems that enable searchers to assess their own knowledge using automatically generated questions~\cite{heilman_language_2006,heilman_personalization_2010,syed_improving_2020}. For example, Syed et al.~\cite{syed_improving_2020} evaluated a system that prompted participants to answer automatically generated factoid questions to assess their comprehension of passages read during the session (i.e., remember/factual learning).  Future research could extend such systems by asking more complex questions to support a wider range of learning objectives during search.
%\vspace{-.3cm}
\subsection{Tools to Support Procedural Learning Objectives}\label{subsec:implications_procedural}

Next, in Table~\ref{tab:implications-summary}, we discuss four potential tools and features to support searchers during procedural learning objectives.

Our results found that procedural objectives involved more create LIs (RQ2) and create-to-create transitions (RQ3).  As previously noted, during procedural objectives, participants often simplified, modified, combined, and created new procedures based on their unique preferences (e.g., familiar techniques) and constraints (e.g., available tools and materials).  We propose three tools to support such create-level processes during procedural objectives.  

First, systems should prioritize \emph{linking} procedures through ``querying-by-example'', enabling searchers to efficiently find alternative procedures for the same task.  Additionally, systems could display related procedures side-by-side and emphasize their similarities and differences.  Helping searchers find and compare related procedures may enable them to: (1) gauge the range of alternative approaches to the task at hand; (2) identify steps that are skippable or modifiable; and (3) discover different ways to execute a step.  Ultimately, such tools could help searchers \emph{combine} ideas from related procedures to fit their preferences and constraints.

Second, during procedural objectives, create LIs often occurred when participants considered alternative techniques, tools, and materials to implement a specific procedure.  To support this process, search systems could automatically identify and suggest alternative techniques, tools, and materials when a searcher is reviewing a specific procedure.  For example, if a recipe calls for heavy cream, the system could automatically point out that half-and-half is a common replacement.\footnote{Procedural knowledge sites such as Wikihow often include user-generated comments that mention functionally equivalent techniques, tools, and materials. A system could mine such resources to identify interchangeable techniques, tools, and materials.}

Third, participants sometimes used concepts (and their definitions) as inspiration to create new procedures.  For example, one participant used the concept of a diameter to develop their own approach for finding the center of a circle.\footnote{Use a string to find the longest ``chord'' of the circle (i.e., the diameter), and then fold the string in half to find the center of the circle.} To better support this process, search systems could display related concepts in response to a procedural knowledge query.

Our results also found that procedural objectives had more upshift transitions to more complex processes (RQ3).  To support these upshifts, systems could enable searchers to ``query-by-example'' based on explicitly stated goals.  In other words, the system could enable searchers to submit a specific procedure as a ``query'' and explicitly request either: (1) background information (i.e., to support upshifts to understand); (2) example videos of the procedure being executed (i.e., to support upshifts to apply); (3) alternative procedures with the same objectives (i.e., to support  upshifts to analyze); (4) pros and cons of the procedure (i.e., to support upshifts to evaluate); or (3) potential modifications (i.e., to support upshifts to create).

%\vspace{-.3cm}
\subsection{Tools to Support All Learning Objectives}\label{subsec:implications_all}

Finally, in Table~\ref{tab:implications-summary}, we discuss four potential tools and features to support searchers with learning objectives of any knowledge type.

First, our results found that remember and understand were frequent LIs regardless of the objective (RQ2).  In other words, as might be expected, remember- and understand-level processes seem to be foundational in support of more complex subgoals.  To support searchers with remember- and understand-level processes, systems should provide interactive spaces for searchers to copy/paste, summarize, organize, and annotate information as they search.  Prior studies have found that such tools can provide learning benefits~\cite{qiu_towards_2020,roy_note_2021,freund_effects_2016}.  As an implication for future research, we believe that such tools should be designed to support and encourage both remember- \emph{and} understand-level processes.  Remember-level processes can be easily supported by allowing searchers to copy/paste and save information.  More importantly, to support understand-level processes, tools should encourage searchers to summarize information in their own words and organize information using their own knowledge representations (e.g., labeled clusters).

Second, downshifts to understand were common regardless of the learning objective (RQ3).  To support these downshifts, systems could provide users with search trails of their own search history.  In our study, participants often downshifted to understand by revisiting pages to reread content or reissuing queries to examine alternative search results.  Providing users with access to their own search trails could streamline this process.  Prior studies have investigated the benefits of search trails from \emph{other} searchers who performed a similar task~\cite{ capra_differences_2015,White2010}.  Our results suggest that users may also benefit from having access to their \emph{own} search trails during a learning-oriented session.

Third, transitions back to the same cognitive process were frequent regardless of the learning objective (RQ3).  For example, if a searcher is evaluating, it is likely that they will continue to evaluate in the next subgoal.  We noticed this happening when participants wanted: (1) simpler or more complex content to better \emph{integrate} new information with their existing knowledge; (2) content from a different source to \emph{verify} newly acquired knowledge; or (3) content from a different perspective to \emph{deepen} their knowledge.  This trend is consistent with results from Liu et al.~\cite{liu2020identifying}, which found that learning-oriented searches often involve repeated iterations of the same overall intent. To support these iterative transitions, systems should enable users to filter search results along dimensions such as complexity (or target audience), originating source, and perspective. Developing systems that can filter results by complexity and source seems relatively straightforward. Developing systems that can filter results by perspective seems more challenging and is area of ongoing research. Tabrizi et al.~\cite{Tabrizi2018} describe the different hypothetical components of a perspective-based search system.

Finally, upshifts from understand to analyze were frequent regardless of the learning objective (RQ3). This means that understanding a fact, concept, or procedure, typically leads to analyzing its relation (e.g., similarities and differences) to other facts, concepts, and procedures.  In other words, understanding something in isolation typically leads to understanding it in a greater framework. To support these understand-to-analyze transitions, systems should enable searchers to explore related facts, concepts, and procedures in response to an understand-level query.  Commercial systems already do this to some extent (albeit inconsistently). For example, in response to the query ``burj khalifa skyscraper height'', Google also shows results for other skyscrapers in the ``People also search for'' section. By making this related information more accessible, searchers may be able to situate newly acquired knowledge in a greater framework.  For example, the Burj Khalifa is 2,717 feet tall and (more impressively) more than twice as tall as the Empire State Building. 

As an implication for future work, systems should enable searchers to explore related facts, concepts, and procedures in a more \emph{consistent} and \emph{self-directed} manner (e.g., ``See related [facts | concepts | procedures]''). Enabling searchers to explore related facts, concepts, and procedures may encourage searchers to transition from understand- to analyze-level processes and may provide different benefits.  As in the example above, exploring related facts may help searchers gain a deeper appreciation of a fact. Exploring related concepts may help searchers understand the unique aspects of a concept in relation to similar concepts.  Exploring related procedures (i.e., procedures to accomplish a different but related task) may help searchers deepen their understanding of important concepts and techniques used in related tasks.  Ultimately, such tools could help searchers situate newly acquired knowledge in a greater framework (i.e., an analyze-level process).

\newgeometry{left=1.5cm,top=2cm,right=1.5cm}
\thispagestyle{empty}
\begin{table}[]
\renewcommand\thetable{15}
\vspace{-0.2cm}
\caption{\small Implications for search system modifications based on findings (section~\ref{sec:results}).}
\vspace{-0.2cm}
\label{tab:implications-summary}
\footnotesize
\begin{tabularx}{\textwidth}{|p{4cm}|p{4.5cm}|p{4.5cm}|p{4.5cm}|}
\hline
Learning Objective Type & Potential Modification/Tool & Results Supported & Related Work 
\\ \hline
 & \textbf{Organize results by definitions, overviews, and examples} & 
 \begin{itemize}[leftmargin=8pt, labelindent=0pt, itemindent=0pt] 
    \item Conceptual LOs involved more understand (RQ2)
 \end{itemize} &  \\ \cline{2-4} 
 & \textbf{Organize examples by type} & 
 \begin{itemize}[leftmargin=8pt, labelindent=0pt, itemindent=0pt] 
    \item Conceptual LOs involve downshifts to understand (RQ3)
 \end{itemize} &  \\ \cline{2-4} 
\multirow{-3}{*}{\parbox{3cm}{\textit{Conceptual Learning \\ Objectives}}} & \textbf{Questions posed to test understanding; e.g., ``Which of these images exemplify Bernoulli's principle?''} & \begin{itemize}[leftmargin=8pt, labelindent=0pt, itemindent=0pt]         \item Conceptual LOs involve downshifts to understand (RQ3)        \end{itemize} & 
  \begin{itemize}[leftmargin=8pt, labelindent=0pt, itemindent=0pt]       \item Question generation~\cite{syed_improving_2020} 
  \end{itemize}  \\ \hline
 & \textbf{Linking procedures through ``querying-by-example'' and related procedures side-by-side} & 
 \begin{itemize}[topsep=0pt, leftmargin=8pt, labelindent=0pt, itemindent=0pt]
    \item Procedural LOs involved more creating (RQ1, RQ2)
    \item Procedural LOs involve upshifts (RQ3)
 \end{itemize} &  \\ \cline{2-4}
 & \textbf{Substitute or functionally equivalent materials} & \begin{itemize}[topsep=0pt, leftmargin=8pt, labelindent=0pt, itemindent=0pt] 
    \item Procedural LOs involved more creating (RQ1, RQ2)
 \end{itemize} &  \\ \cline{2-4}
 \multirow{-4}{*}{\parbox{3cm}{\textit{Procedural Learning \\ Objectives}}} & \textbf{Relevant conceptual knowledge to inspire the creation of novel procedures} & 
 \begin{itemize}[topsep=0pt, leftmargin=8pt, labelindent=0pt, itemindent=0pt] 
    \item Procedural LOs involved more creating (RQ1, RQ2)
 \end{itemize} &  \\ \hline
  & \textbf{Interactive space to copy, summarize, organize, connect, and modify evolving knowledge} 
 & \begin{itemize}[leftmargin=8pt, labelindent=0pt, itemindent=0pt]
     \item Downshifts to understand common in all LOs (RQ3)
     \item Factual LOs involved more remember (RQ2)
     \item Procedural LOs involved more creating (RQ1, RQ2)
 \end{itemize}
 & \begin{itemize}[leftmargin=8pt, labelindent=0pt, itemindent=0pt] 
    \item \raggedright Sticky-note tool~\cite{freund_effects_2016}
    \item Note-taking~\cite{qiu_towards_2020} 
    \item Note-taking and highlighting~\cite{roy_note_2021} 
   \end{itemize}
 \\ \cline{2-4}
 & \textbf{Trails of searcher's own search history} & 
 \begin{itemize}[leftmargin=8pt, labelindent=0pt, itemindent=0pt]
    \item Downshifts to understand common in all LOs (RQ3)
 \end{itemize}
 & \begin{itemize}[leftmargin=8pt, labelindent=0pt, itemindent=0pt] 
    \item \raggedright Search trails of other searchers~\cite{ capra_differences_2015,White2010} 
   \end{itemize} \\ \cline{2-4} 
 & \textbf{Filter search results across dimensions such as complexity, perspective, and originating source}                  
 & \begin{itemize}[leftmargin=8pt, labelindent=0pt, itemindent=0pt]       \item Transitions back to the same cognitive process common in all LOs (RQ3)
 \end{itemize} & \begin{itemize}[leftmargin=8pt, labelindent=0pt, itemindent=0pt]       \item Perspective-based search system~\cite{Tabrizi2018}
 \end{itemize} \\ \cline{2-4} 
\multirow{-5}{*}{\textit{All Learning Objectives}} & \textbf{Related facts, concepts, and procedures; highlight differences/similarities} & \begin{itemize}[leftmargin=8pt, labelindent=0pt, itemindent=0pt] 
    \item Understand to analyze common in all LOs (RQ3)
    \item Factual LOs involve downshifting to remember (RQ3)
    \item Conceptual LOs involve downshifting from evaluate to analyze (RQ3)
\end{itemize} &  \\ \hline
\end{tabularx}
\end{table}
\restoregeometry

\section{Caveats and Opportunities for Future Work}

Our study and results have a few caveats worth noting.

First, participants completed tasks with learning objectives associated with three (out of six) cognitive processes from A\&K's taxonomy. As previously mentioned, we omitted the cognitive processes of remember and understand because they are the least complex, and we omitted analyze because analyzing is a necessary component of evaluating. Future work might consider the full range of cognitive processes as \emph{learning objectives}.

Second, participants were given a maximum of 15 minutes to complete the search phase of each task.  Naturally, time constraints can influence search behaviors and outcomes~\cite{Crescenzi2021}.  In our case, the 15-minute time limit may have influenced the pathways taken by participants to achieve the given objective.  Future work is needed to investigate how time constraints may influence the pathways taken by searchers towards an objective. In our study, the 15-minute time limit was imposed to keep the study session under 1.5 hours.  Additionally, after several rounds of pilot testing, we determined that 15 minutes was enough time for participants to complete our tasks. Ultimately, participants spent about 10 minutes searching on average ($M=9.79$, $S.D.=5.39$).\footnote{In Urgo et al.~\cite{UrgoICTIR2020}, we reported on the effects of the objective on task completion time.  The cognitive process of the objective had no significant effects.  Conversely, participants took longer to complete tasks with conceptual objectives (11.65 minutes) versus procedural objectives (8.43 minutes).}

% The median was $MD=10.79$.  Not sure if we want to mention it.

Finally, in our study, we did not explicitly consider knowledge gain as a main dependent variable. In other words, participants did not complete pre- and post-tests to measure learning during the search process. Instead, participants were asked to demonstrate their achievement of the learning objective in a 2-minute video recorded by the moderator. We believe that the video demonstration phase of each task encouraged participants to achieve the given objective. Participants had to explain their main solution to the task ``live'' and in front of the moderator.

In terms of learning outcomes, we see important opportunities for future work. In our study, we used qualitative techniques to investigate the learning \emph{process} during search. Future studies could adopt similar techniques to investigate how the learning process might influence learning \emph{outcomes}.\footnote{In Section~\ref{sec:related_work}, we briefly summarize different methods used to quantify knowledge gains during the search process.}  An important open question is: What are characteristics of pathways that lead to greater knowledge gains?  For example, are greater knowledge gains achieved when pathways are longer, more diverse, or have more upshift transitions? Our study focused on how objectives influence pathways. Future work should also consider how pathways impact learning.

\section{Conclusion}\label{sec:conclusion}

In this paper, we have introduced the notion of a pathway---a sequence of learning instances (or subgoals) followed by a searcher towards a specific learning objective. We leveraged A\&K's taxonomy~\cite{anderson2001taxonomy} to characterize both learning objectives and pathways. We studied the impact of a objective's knowledge type and cognitive process on the pathway length, diversity, and cognitive processes traversed. Additionally, we analyzed the transition probabilities between cognitive processes conditioned on the knowledge type of the objective.

Our research makes several important contributions. First, from a methodological perspective, we present a method for analyzing search sessions to gain insights about the learning \emph{process} during search.  In this paper, we analyzed how the learning objective (i.e., the end goal) can influence the learning process.  Future studies could use our methodology to more closely investigate how pathway characteristics influence learning outcomes. 

Second, our results found that the objective can influence the pathways followed by searchers.  Importantly, the knowledge type of the objective had a much stronger effect than the cognitive process.  For example, factual objectives involved more \emph{remembering} (e.g., memorizing), conceptual objectives involved more \emph{understanding} (e.g., summarizing and exemplifying), and procedural objectives involved more \emph{creating} (e.g., simplifying, modifying, and combining).  Prior studies have also leveraged A\&K's taxonomy to investigate how objectives impact search behaviors and outcomes. However, studies have primarily leveraged the cognitive process dimension and ignored the knowledge type dimension.  Our results indicate that searchers may need different types of support during objectives involving factual, conceptual, and procedural knowledge.

Finally, our analysis of cognitive process transitions revealed several important trends. Our results found several likely transitions irrespective of the objective. Participants were likely to start with understand, iterate on the same cognitive process, transition from high- to low-complexity processes, and transition from understand to analyze. Additionally, our results found several transitions to be likely \emph{depending} on the knowledge type of the objective. For example, during factual objectives, participants were more likely to start with remember and transition to remember from more complex processes. During conceptual objectives, participants were more likely to start with understand and transition \emph{back} to understand from more complex processes. During procedural objectives, participants were the most likely to transition \emph{upwards} from low- to high-complexity processes. We have discussed implications from our results for designing search environments (i.e., search features and auxiliary tools) to support learning. In future work, we plan to explore the impact of \emph{scaffolding} tools that support the development and externalization of subgoals based on the pathways observed in our study.

\bibliographystyle{unsrt}  
\bibliography{references}  %%% Remove comment to use the external .bib file (using bibtex).


% George Kour and Raid Saabne.
% \newblock Real-time segmentation of on-line handwritten arabic script.
% \newblock In {\em Frontiers in Handwriting Recognition (ICFHR), 2014 14th
%   International Conference on}, pages 417--422. IEEE, 2014.

% George Kour and Raid Saabne.
% \newblock Fast classification of handwritten on-line arabic characters.
% \newblock In {\em Soft Computing and Pattern Recognition (SoCPaR), 2014 6th
%   International Conference of}, pages 312--318. IEEE, 2014.

% \bibitem{hadash2018estimate}
% Guy Hadash, Einat Kermany, Boaz Carmeli, Ofer Lavi, George Kour, and Alon
% \newblock Estimate and replace: A novel approach to integrating deep neural
%   networks with existing applications.
% \newblock {\em arXiv preprint arXiv:1804.09028}, 2018.

% \end{thebibliography}

%\pagebreak

\appendix
\begin{appendices}
\section{Appendix}

\subsection{Coding Guide}\label{subsec:coding_guide}
\begin{center}
\rule{\textwidth}{0.02cm}
\end{center}

The coding guide was developed based on A\&K's book~\cite{anderson2001taxonomy}, situating learning objectives at the intersection of two orthogonal dimensions: the cognitive process dimension and the knowledge type dimension. This coding guide associates learning instances (LIs) with a particular cognitive process \emph{and} a particular knowledge type. Because metacognitive knowledge was not part of our analysis, the guide includes 3 knowledge types (factual, conceptual, and procedural) and 6 cognitive processes (remember, understand, apply, analyze, evaluate, and create). Thus, there are 18 possible coding combinations. Each of these combinations (e.g., Factual/Remember) are outlined below.

The LI units consisted of recorded screen activities (e.g., queries) and think-aloud comments. When considering how to categorize an LI, the coder asked themselves “What is the intention of the learner at this moment?” Then, the coder used the following criteria to determine the most appropriate category. Criteria for these categories include particular actions and associated examples.

\vspace{2mm}
\hrule
\vspace{0.5mm}
\hrule
\vspace{2mm}
\textbf{Factual}
\vspace{2mm}
\hrule
\vspace{0.5mm}
\hrule
\vspace{2mm}
\emph{Factual/Remember}
\begin{itemize}
 \item Reads a fact (e.g., person, date, place)
 \item Copy/pastes a fact without additional verbal or written elaboration
 \item Memorizes a fact
 \item Issues a query using specific language of fact
 \begin{itemize}
 \item e.g., “tallest building”
 \end{itemize}
 \item Notes having seen or heard of fact before
\end{itemize}

\emph{Factual/Understand}
\begin{itemize}
 \item Finds example of fact in a visualization
 \item Restates fact in own words
 \item Makes an estimation of a fact
 \item Writes down a fact with additional verbal elaboration
 \item Acknowledges or includes source of fact
 \item Makes a query using general language of fact
 \begin{itemize}
 \item e.g., “tall buildings”
 \end{itemize}
 \item Investigates a particular fact beyond just noting/reading the fact
 \begin{itemize}
 \item e.g., “I don't know Abraham Lincoln. Let's look into him a bit more.”
 \item e.g., “This artist keeps coming up, so I'll look into them a bit more.”
 \end{itemize}
 \item Makes some inference about fact
 \begin{itemize}
 \item e.g., “This mentions that Americans have an average of 2.5 doctors for every 1,000 people while Europeans have an average of 3.5, we Americans must generally be less healthy.” 
 \end{itemize}
\end{itemize}

\emph{Factual/Apply}
\begin{itemize}
 \item Uses one fact with second fact to find third fact
\end{itemize}

\emph{Factual/Analyze}
\begin{itemize}
 \item Compares and differentiates facts that clarify or explain fact
 \item Compares two facts to arrive at most reasonable average or estimation
 \item Differentiates between categories of facts
 \begin{itemize}
 \item e.g., “Oh I see this is the number of years a cat lives, but this is the number of human years a cat lives.”
 \end{itemize} 
 \item Compares facts to find correlations or themes among the facts or generalizations
 \begin{itemize}
 \item e.g., “Most of the deep parts of the ocean are in the Pacific Ocean.”
 \end{itemize} 
 \item Lists or identifies correlations (or lack of correlation) or themes among the facts
 \item Identifies an outlier among correlations or themes
 \begin{itemize}
 \item e.g., “Most of these are in the Pacific Ocean, but I see this point is in the Atlantic.”
 \end{itemize}
 \item Disambiguates two facts
 \begin{itemize}
 \item e.g., “The Challenger Deep exists within the Mariana Trench, they are not separate locations.”
 \end{itemize}
\end{itemize}
\vspace{2mm}

\emph{Factual/Evaluate}
\vspace{1mm}

The existence of a “judgment” word or phrase does not in and of itself elicit an evaluate code (e.g, “that's interesting”, “so that's good”, or “that's not what we need” are not coded as evaluate). It is important that the judgment is made with respect to the learning subgoal or overall objective (e.g., “I can see that most of these paintings are from post-1900, so date of creation is a good potential explanation for the high cost of the painting.”)
\begin{itemize}
 \item Judges, critiques, or questions validity of a fact
 \item Judges or critiques importance or validity of a fact that explains other fact
 \item Judges, critiques, or questions resource validity of fact
 \item Verifies a previously known fact.
 \item Making a judgment about a fact being good/bad at explaining something.
 \item States why a fact is good/bad at explaining something.
 \item Judges usefulness of criteria for assessing reasoning behind fact
 \begin{itemize}
 \item e.g., “This cat breed chart's variables are a good way to assess why cats are such popular pets.”
 \end{itemize}
 \item Selects a reason for explanation of fact
 \begin{itemize}
 \item e.g., “I think this cat is so expensive because the breed is so rare.”
 \end{itemize}
 \item Compares lists of facts to find agreement
 \begin{itemize}
 \item e.g., “I'm actually not finding a lot of overlap between the two lists of most expensive paintings.”
 \end{itemize}
\end{itemize}

\emph{Factual/Create}
\begin{itemize}
 \item Generates new logical reason substantiating a fact that is not explicitly or implicitly cited in resource
\end{itemize}

\vspace{2mm}
\hrule
\vspace{0.5mm}
\hrule
\vspace{2mm}
\textbf{Conceptual}
\vspace{2mm}
\hrule
\vspace{0.5mm}
\hrule
\vspace{2mm}
%Note: If a fact is considered in an effort to better understand a concept then it is coded as conceptual.
\vspace{2mm}

\emph{Conceptual/Remember}
\begin{itemize}
 \item Reads through definition of concept
 \item Memorizes definition of concept
 \item Writes down or copy/pastes definition of concept without additional verbal or written elaboration
 \item Reads through (notes) techniques or concepts associated with main concept without additional verbal or written elaboration
 \item Reads through (notes) definitions of techniques or concepts associated with main concept without additional verbal or written elaboration
 \item Reads or notes (not queries) name of concept without additional verbal or written elaboration
 \item Recalls definition of concept
\end{itemize}

Note: If examples are involved in \emph{any} way then move up (at least) to understand
\vspace{2mm}

\emph{Conceptual/Understand}
\begin{itemize}
 \item Reviews example of concept (e.g., reads the labeled parts of an example of concept, looks at paintings that exemplify an artistic style)
 \item Summarizes definition of concept in own words
 \item Writes down concept and elaborates verbally
 \begin{itemize}
 \item e.g., notes concept and states that they don't understand the concept
 \item e.g., notes concept and states they will look into it later
 \end{itemize}
 \item Reviews examples of techniques or concepts associated with main concept
 \item Summarizes definition of techniques or concepts associated with main concept
 \item Reviews example or representative work of concept
 \item Reviews example or representative work of technique associated with main concept
 \item States that they are attempting to “understand” a technique or concept
 \item Makes a guess at the definition of a concept in own words
 \begin{itemize}
 \item e.g., “So it seems like surrealism might be art that is made from an artist's perception of dreams or other alternate realities.”
 \end{itemize} 
 \item Makes some inference about concept
 \begin{itemize}
 \item e.g., “It says that insurance can be included in a mortgage, so it's probably not mandatory.”
 \end{itemize} 
 \item Issues exploratory query of concept
 \begin{itemize}
 \item e.g., “‘I really don't know what this is, but I'll start with this.' bernoulli's principle”
 \end{itemize} 
 \item Reads, writes down or notes example of concept
 \begin{itemize}
 \item e.g., “So water through a hose with a thumb partially covering the hole is an example of Bernoulli's principle.”
 \end{itemize} 
 \item Recognizes explanations of phenomena that use different concepts
 \begin{itemize}
 \item e.g., “So there are two different explanations for why the sky is blue, Tyndall Effect and Rayleigh Scattering.” 
 \end{itemize}
 \item Uncovers conflicting or disputed information
 \begin{itemize}
 \item e.g., “So it looks like there is disagreement on which is the appropriate theory to explain why the sky is blue.”
 \end{itemize} 
\end{itemize}

\emph{Conceptual/Apply}
\begin{itemize}
 \item Talks through application of concept (e.g., expresses how essential components of concept function or are represented in a given example)
 \item Questions whether example is instance of concept by citing definition or characteristics of concept
 \item Applies essential components of concept to specific novel example
 \item Revisits example to apply participant's newly acquired characteristics of concept to example
 \begin{itemize}
 \item e.g., “Ah yes, I see now that this is made up of corners consisting of perpendicular lines as I learned from the definition of a square.”
 \end{itemize}
 \item Investigates a particular application of a concept to a scenario
 \begin{itemize}
 \item e.g., “I understand the general idea of Newton's second law, but I am really interested in how this applies to a kayak moving through the water.”
 \end{itemize}
 \item Applies concept to example from memory (example is not explicitly stated in site)
 \begin{itemize}
 \item e.g., “oh so the Tyndall Effect is like when you see blue smoke from motorcycle exhaust.” (when motorcycle example had not ever been mentioned during search session)
 \end{itemize}
\end{itemize}

\emph{Conceptual/Analyze}
\begin{itemize}
 \item Differentiates between two or more concepts
 \item Connects the dots among examples, extrapolates, generalizes concept
 \begin{itemize}
 \item e.g., “So it looks like all of these are abstract and involve lions.”
 \item e.g., “The image search is not really bringing up what I expected...it's almost like it's [concept is] broader than I assumed.” 
 \end{itemize}
 \item Differentiates between forms, representations, or types of concept
 \begin{itemize}
 \item e.g., “If this needs to include all systems of democracy I should include parliamentary and presidential.” 
 \end{itemize}
 \item Groups examples by form, representation or type of concept
 \item Realizes that two different concepts together best explain a phenomena
 \begin{itemize}
 \item e.g., “Ok, so I guess you need to know about the continuity equation and Bernoulli's equation to analyze the flow of a fluid.”
 \end{itemize}
\end{itemize}

%Example sequence: looks at a bunch of examples to understand concept, higher level overview of characteristics/components of concept to see similarities/differences among examples (analyze), and then moves on to other example to apply characteristics/components of concept to new example
\vspace{2mm}

\emph{Conceptual/Evaluate}
\vspace{1mm}

The existence of a “judgment” word or phrase does not in and of itself elicit an evaluate code (e.g, “that's interesting”, “so that's good”, or “that's not what we need” are not coded as evaluate). It is important that the judgment is made with respect to the learning subgoal or overall objective (e.g., “I think Bernoulli's principle is easier to understand than Newton's laws of motion when explaining lift.”)
\begin{itemize}
 \item Judges or critiques effectiveness, appropriateness, and/or validity of concept to explain a phenomenon
 \item Judges or critiques a concept given some characteristics (e.g., efficiency, effectiveness, context, complexity, ease, appropriateness)
 \begin{itemize}
 \item e.g., “So that makes sense and it's a little bit easier than the general definition of Bernoulli's principle.”
 \end{itemize}
 \item Judges or critiques usage of concept
 \item Judges, critiques, or questions resource validity of concept
 \begin{itemize}
 \item e.g., “In order to be super thorough I am going to visit another link just to make sure the information about the components of a mortgage are consistent across sources.”
 \end{itemize}
\end{itemize}

\emph{Conceptual/Create}
\begin{itemize}
 \item Generates new representation of concept
 \item Draws connections among concepts beyond those explicitly or implicitly stated in resource
\end{itemize}

\vspace{2mm}
\hrule
\vspace{0.5mm}
\hrule
\vspace{2mm}
\textbf{Procedural}
\vspace{2mm}
\hrule
\vspace{0.5mm}
\hrule
\vspace{2mm}
\emph{Procedural/Remember}
\begin{itemize}
 \item Reads or notes steps of procedure
 \item Rotely memorizes steps of procedure
 \item Copy/pastes steps of procedure
\end{itemize}

\emph{Procedural/Understand}
\begin{itemize}
 \item Summarizes steps of procedure
 \item Reviews example of procedure with verbal or written elaboration
 \item Reviews diagram of procedure's end product with verbal or written elaboration
 \item Issues exploratory query of procedure
 \begin{itemize}
 \item e.g., ways of paper airplane making
 \end{itemize}
 \item Notes some characteristic of procedure
 \begin{itemize}
 \item e.g., “So that said it would be x number of steps but it is really y number of steps.”
 \end{itemize}
\end{itemize}

\emph{Procedural/Apply}
\begin{itemize}
 \item Executes/narrates/visualizes a procedure
\end{itemize}

\emph{Procedural/Analyze}
\begin{itemize}
 \item Explores options (e.g., materials, steps, tools) that meet criteria for procedure
 \item Compares and contrasts two or more procedures
 \begin{itemize}
 \item e.g., “This one has a lot of steps, this one has fewer steps. It seems like there is a wide range of steps that it might take to fold an airplane.”
 \end{itemize}
 \item Selects essential components of procedure
 \item Explores tradeoffs of a particular method or methods
 \item Explores suggestions or techniques for improving execution of procedure
 \item Compares own version of steps of procedure with procedure on site
 \begin{itemize}
 \item e.g., “I'm just going to make sure my steps that I've written down match with what I see.”
 \end{itemize}
 \item Identifies important techniques in executing procedure that make the end product most effective
 \begin{itemize}
 \item e.g., “I think it is important to hold the string very taught to make sure it is a straight line, and it is also important to make sure both lengths of string are measured exactly or it won't be accurate.”
 \end{itemize}
 \item Assessing which of the steps need to be executed for personal subjective context
 \begin{itemize}
 \item e.g., “I already have a starter for my bread dough, so I won't need to do the first three steps; I can start with step four.”
 \end{itemize}
\end{itemize}
\vspace{2mm}

\emph{Procedural/Evaluate}
\vspace{1mm}

The existence of a “judgment” word or phrase does not in and of itself elicit an evaluate code (e.g, “that's interesting”, “so that's good”, or “that's not what we need” are not coded as evaluate). It is important that the judgment is made with respect to the learning subgoal or overall objective (e.g., “I think the Kite would be a better choice for my nephew because it can fly farther and do more tricks than the Hammer.”)
\begin{itemize}
 \item Judges or critiques a procedure given some characteristics (e.g., efficiency, effectiveness, context, complexity, expertise of executor, ease, enjoyment, attractiveness, appropriateness )
 \item Checks that procedure was executed correctly
 \item Outlines criteria used to judge methods
 \begin{itemize}
 \item e.g., “So I am looking for methods that are challenging, interesting, and have many steps.”
 \end{itemize}
 \item Makes judgement that a procedure is difficult or too difficult/not worth the effort
 \item Makes judgement that a procedure is interesting or fun
 \item Judges steps of procedure to be (in)correct or expressed (in)correctly
 \begin{itemize}
 \item e.g., “It doesn't make sense to do these steps in this order, I wouldn't arrive at the end product.”
 \end{itemize}
 \item Judges, critiques, or questions resource validity of procedure
\end{itemize}

\emph{Procedural/Create}
\begin{itemize}
 \item Finds useful/appropriate replacement materials for a procedure
 \item Modifies step(s) of a procedure in order to suit a particular application
 \item Combines steps of procedures into new procedure
 \item Generates a novel procedure
\end{itemize}

\end{appendices}

\end{document}